\documentclass[11pt]{article}
\usepackage{graphicx} 
\usepackage{txfonts}

\begin{document} 
   
\title{Magnetic Field Evolution in Neutron Stars: One-Dimensional Multi-Fluid Model}

\author{J. Hoyos$^*$, A. Reisenegger$^*$, J. A. Valdivia$^+$ \vspace{0.5cm}\\
$^*$Departamento de Astronom\'ia y Astrof\'isica\\ Pontificia Universidad Cat\'olica de Chile, Santiago, Chile\\
$^*$jhoyos@astro.puc.cl, $^*$areisene@astro.puc.cl \and $^+$Departamento de F\'isica, Facultad de Ciencias, Universidad de Chile, \\
Santiago, Chile\\$^+$alejo@macul.ciencias.uchile.cl} 
  
\maketitle
     
\begin{abstract} 
This paper is the first in a series that aims to understand the long-term evolution of neutron star magnetic fields. We model the stellar matter as an electrically neutral and lightly-ionized plasma composed of three moving particle species: neutrons, protons, and electrons; these species can be converted into each other by weak interactions (beta decays), suffer binary collisions, and be affected by each other's macroscopic electromagnetic fields. Since the evolution of the magnetic field occurs over thousands of years or more, compared to dynamical timescales (sound and Alfv\'en) of milliseconds to seconds, we use a slow-motion approximation in which we neglect the inertial terms in the equations of motion for the particles. This approximation leads to three nonlinear partial-differential equations describing the evolution of the magnetic field, as well as the movement of two fluids: the charged particles (protons and electrons) and the neutrons. These equations are first rather than second order in time (involving the velocities of the three species but not their accelerations). In this paper, we restrict ourselves to a one-dimensional geometry in which the magnetic field points in one Cartesian direction, but varies only along an orthogonal direction. We study the evolution of the system in three different ways: (i) estimating timescales directly from the equations, guided by physical intuition; (ii) a normal-mode analysis in the limit of a nearly uniform system; and (iii) a finite-difference numerical integration of the full set of nonlinear partial-differential equations. We find good agreement between our analytical normal-mode solutions and the numerical simulations. We show that the magnetic field and the particles evolve through successive quasi-equilibrium states, on timescales that can be understood by physical arguments. Depending on parameter values, the magnetic field can evolve by ohmic diffusion or by ambipolar diffusion, the latter being limited either by interparticle collisions or by relaxation to chemical quasi-equilibrium through beta decays. The numerical simulations are further validated by verifying that they satisfy the known conservation laws in highly nonlinear situations. 
\end{abstract}

\section{Introduction}
\label{intro} 
 
Observations of the surface magnetic fields of neutron stars have shown a correlation between the age of the star and the magnetic-field strength. For instance, young radio pulsars and high-mass X-ray binaries have surface magnetic fields of the order of $10^{12}$ G, while millisecond pulsars and low-mass X-ray binaries, which are older objects, have magnetic fields $\sim 10^{8}$ G. These observations suggest that the magnetic field is decaying, although this might be a by-product of accretion (\cite{B-95}; \cite{CZB-01}; \cite{C-02}; \cite{PM-04}). On the other hand, it is thought that the spontaneous decay of the ultrastrong $10^{14-15}$ G magnetic field in \emph{magnetars} is the main source of their X-ray luminosity since these objects appear to radiate substantially more power than that available from their rotational energy loss (\cite{DT-92}; \cite{TD-96}; \cite{ACT-04}). Since these objects appear to be isolated, the field decay must be attributed to processes intrinsic to the neutron stars. 
 
Ref.\cite{GR-92}; (hereafter GR-92) identified three long-term mechanisms that can promote the spontaneous decay of the magnetic field in the interior of neutron stars: 
\begin{enumerate} 
\item  \emph{Ambipolar diffusion}, i.e. the drift of the magnetic field and the charged particles relative to the neutrons. Ref. \cite{TD-96}; (hereafter TD-96), using magnetar parameters, estimated the timescale of the magnetic field decay by means of ambipolar diffusion, finding that was consistent with the timescale of $10^{3-5}$ years observed for these objects. 
\item \emph{Hall drift}, a non-dissipative advection of the magnetic field by the associated electrical current (\cite{J-88}; \cite{RG-02}; \cite{GR-02}; \cite{HR-02}, \cite{HR-04}; \cite{RKG-04}; \cite{CAZ-04}; \cite{R-05}, \cite{R-07}; \cite{Rei-07}). 
\item \emph{Ohmic diffusion}, a dissipative process caused by electrical resistivity (see also Baym et al. \cite{Bay-69}). 
\end{enumerate} 
All of these processes occur over timescales of thousands of years or more, compared to typical dynamical timescales (sound and Alfv\'en) of miliseconds to seconds. The work of GR-92 was analytical, and therefore useful to identify general processes and relevant timescales, but it did not address the action of the identified processes in their full nonlinear development and their interactions with each other. The full evolution of the magnetic field can only be addressed by numerical simulations. 
 
Regarding the geometrical configuration of the magnetic field, three-dimensional magnetohydrodynamics (MHD) simulations showed that in a stable stratified, non-rotating star, the initial field evolves on a short, Alfv\'en-like timescale to a large-scale equilibrium configuration that can be axisymmetric (\cite{BS-04},~\cite{BS-06}; \cite{BN-06}) or non-axisymmetric (\cite{B-08}), depending on the radial dependence of the initial magnetic-field strength. These simulations, being focused on the evolution of the magnetic field over short dynamical timescales, were carried out within the framework of the MHD theory, which considers the stellar matter to be a single fluid. The evolution of these configurations has not yet been studied over longer timescales, on which relative motions of different particle species are present and the processes studied by GR-92 become important. The description of these long-term processes requires a multi-fluid theory. 
 
In this paper we extend the model of GR-92 by allowing both neutron and charged-particle movements, rather than considering the neutrons as a fixed background (see Sect.~\ref{phys}). For simplicity, we restrict ourselves to a plane-parallel system where the magnetic field points in one Cartesian direction but varies only along an orthogonal direction. Although this one-dimensional model is unrealistic, it allows us to study the basic evolutionary processes in an analytical and numerical fashion with no serious complications before considering a realistic stellar geometry in more dimensions. GR-92 considered all the species in the star as non-interacting fermions; in contrast, we use equations of state including interactions among protons and neutrons. Since in neutron stars the magnetic pressure is much less than the degeneracy pressure of the particles, we treat the particle densities in terms of small perturbations around a non-magnetized background in full equilibrium. Thus, we obtain a system of equations that is nonlinear with respect to the magnetic field but linear with respect to the particle densities (see Sect.~\ref{1dmodel}). 
 
We will show that the magnetic field and particle species evolve through successive quasi-equilibrium states. We describe the basic physical processes of this evolution and estimate analytically the characteristic timescales required to reach these quasi-equilibrium states. We find that these timescales depend on the values of physical parameters in the system, such as the magnetic-field intensity, the collision rate between the charged particles and the neutrons, the weak interaction rate, and the ohmic resistivity (Sect.~\ref{carsca}). To obtain an analytic solution of the system of partial differential equations, we linearize the equations with respect to the magnetic field and find normal-mode solutions (Sect.~\ref{normmodes}) and obtain analytical approximations for the decay times in Appendix \ref{ApA}. We develop numerical simulations (whose basic algorithm is described in Appendix \ref{ApB}) in the linear regime and compare with the normal-mode results (Sect.~\ref{finline}). In Sect.~\ref{finnoli}, we carry out numerical simulations in the nonlinear regime and verify the conservation laws of our model. Finally in Sect.~\ref{conc}, we summarize the main results of this paper and provide our conclusions. 

\section{General Physical Model} 
\label{phys} 
We model the stellar interior as an electrically-neutral and slightly-ionized plasma composed of three moving-particle species: neutrons $(n)$, protons $(p)$, and electrons $(e)$. This is an extension of the model of GR-92, since these authors considered the neutrons to form a motionless background. Furthermore, we account for strong interactions between neutrons and protons by writing each of their chemical potentials as a function of both of their number densities: $\mu_{n,p}(\vec{r},t)=\mu_{n,p}[n_n(\vec{r},t), n_p(\vec{r},t)]$. We consider the electrons to be an ideal, relativistic Fermi gas, and write their chemical potential as a function only of their own density, $\mu_e(\vec{r},t)=\mu_e(n_e(\vec{r},t))=\sqrt{(m_ec^2)^2+\hbar^2c^2[3\pi^2n_e(\vec{r},t)]^{2/3}}$, with $m_e$ the electron rest mass. The plasma species are described as three fluids coupled by collisions and electromagnetic forces, satisfying the equations of motion (GR-92; Reisenegger et al. \cite{R-05}):

\begin{eqnarray} 
\label{movecond} 
n_i\frac{\mu_i}{c^2}\frac{d\vec{v}_i}{dt}&=&-n_i\vec{\nabla}\mu_i-n_i\frac{\mu_i}{c^2}\vec{\nabla}\psi\nonumber\\
&&{}+n_iq_i(\vec{E}+\frac{\vec{v_i}}{c}\times \vec{B})-\sum_{j\neq i}\gamma_{ij}n_in_j(\vec{v}_i-\vec{v}_j),
\end{eqnarray} 
 
\noindent  where  $\vec{v}_i$ is the mean velocity; $d/dt=\partial/\partial t +(\vec{v}_i\cdot \vec \nabla)$, $\mu_i/c^2$ is the effective mass of each species (which includes corrections due to interactions and relativistic effects; \cite{A-98}); $n_i\vec{\nabla}\mu_i$ is the degeneracy-pressure gradient of the species $i$; $\psi$ is the gravitational potential; $\vec{E}$ and $\vec{B}$ are the electric and magnetic fields; and the last term represents the drag forces due to elastic binary collisions, which damp the relative motions of the different species. The collisional coupling strengths are parametrized by the symmetric matrix $\gamma_{ij}$ that depends generally on the local density and temperature. We ignore the effects of superfluidity or superconductivity, as well as additional particles that might be present in neutron stars. 
 
In the early stages of the neutron-star formation, particles are locked by collisions, which behave as a single fluid. Alfv\'en waves propagate across the star, allowing it to reach a \emph{magnetohydrostatic quasi-equilibrium} state in which all forces acting on a fluid element are closely in balance. Since in this paper we are interested in simulating the evolution over far longer timescales characteristic of the processes of GR-92, we use a slow-motion approximation in which we neglect the acceleration terms on the left-hand side (LHS) of Eq.~(\ref{movecond}) (i.e., $(n_i\mu_i/c^2)~d\vec{v}_i/dt=\vec{0}$). We assume implicitly that all the forces acting on a given particle species balance each other at all times. To some extent, this will happen, in the sense that the relative velocities will take such values that the drag forces balance all the other forces, and these drag forces will decay as the other forces come into balance. However, if Eq.~(\ref{movecond}) is summed over all three particle species, the drag forces cancel and no mechanism remains to ensure that the entire fluid reaches magnetohydrostatic quasi-equilibrium. To create such mechanism, we introduce an artificial friction-like force acting on the neutrons (the most abundant species) of the form $-n_n\alpha \vec{v}_n,$ such that the neutron equation of motion becomes  
 
\begin{equation} 
\label{move2} 
\vec{0}=-n_n\vec{\nabla}\mu_n-n_n \frac{\mu_n}{c^2}\vec{\nabla}\psi-\sum_{j\neq n}\gamma_{nj}n_nn_j(\vec{v}_n-\vec{v}_j)-n_n\alpha \vec{v}_n. 
\end{equation} 
 
\noindent As a result, we obtain the equation of magnetohydrostatic quasi-equilibrium for a fluid element, 
 
\begin{equation} 
\label{MHDmod} 
\vec{0}=-\sum_{i=e,p,n} n_i\vec{\nabla}\mu_i-\sum_{i=e,p,n} n_i \frac{\mu_i}{c^2}\vec{\nabla}\psi+\frac{\vec{J}\times\vec{B}}{c}-n_n\alpha\vec{v}_n, 
\end{equation}  
where $\vec{J}=n_ce(\vec{v_p}-\vec{v_e})=(c/4\pi)(\vec{\nabla}\times\vec{B})$ is the electrical current density. Thus, the system approaches magnetohydrostatic quasi-equilibrium on a timescale controlled by the parameter $\alpha$ (see Sect.~\ref{tmhdcar}). This parameter is chosen in such a way that this timescale is sufficiently long for the numerical code to be able to deal with it (and therefore much longer than the true dynamical timescales [sound and Alfv\'en]), but shorter than the long timescales of interest in this paper.  
 
If we add Eq.~(\ref{movecond}) for electrons and protons, assuming charge neutrality, $n_e=n_p\equiv n_c$, we obtain the diffusion equation for the combined fluid of charged particles,

\begin{equation} 
\label{chargeddiff} 
n_cn_n\gamma_{cn}\vec{v}_A=-n_c\vec{\nabla}\mu_c-n_c\frac{\mu_c}{c^2}\vec{\nabla}\psi+\frac{\vec{J}\times \vec{B}}{c}, 
\end{equation}

\noindent where $\mu_c\equiv \mu_e+\mu_p$, $\gamma_{cn}\equiv \gamma_{en}+\gamma_{pn}$, and we define the \emph{ambipolar diffusion velocity} by: 
 
\begin{equation} 
\label{ambi} 
\vec{v}_A \equiv \frac{\gamma_{pn}(\vec{v}_p-\vec{v}_n)+\gamma_{en}(\vec{v}_e-\vec{v}_n)}{\gamma_{cn}}. 
\end{equation} 
If the charged particles are not in \emph{diffusive quasi-equilibrium}, that is the forces on the right-hand side (RHS) of Eq.~(\ref{chargeddiff}) are not in balance, the latter will drive them to move relative to the neutrons at the ambipolar diffusion velocity $\vec{v}_A$. On the other hand, if the system is in magnetohydrostatic quasi-equilibrium (see Eq.~(\ref{MHDmod}) and the charged particles have reached their diffusive quasi-equilibrium ($\vec{v}_A =0$), this implies that neutrons are also in a diffusive quasi-equilibrium state given by  
 
\begin{equation} 
\vec{0}=-n_n\vec{\nabla}\mu_n-n_n\frac{\mu_n}{c^2}\vec{\nabla}\psi. 
\end{equation} 
 
To derive the equation that governs the evolution of the magnetic field, we combine Eq.~(\ref{movecond}), for electrons and protons without the inertial terms, and the induction equation $\vec{\nabla}\times \vec{E}=-(1/c)(\partial \vec{B}/\partial t)$, to obtain

\begin{eqnarray} 
\label{magn} 
\frac{\partial \vec{B}}{\partial t}&=&\vec{\nabla}\times \left[(\vec{v}_n+\vec{v}_A+\vec{v}_H)\times \vec{B}\right]-\vec{\nabla}\times \left(\frac{c^2\vec{\nabla}\times{\vec{B}}}{4\pi\sigma}\right)\nonumber\\&&
-\frac{c}{2e}\vec{\nabla}\left(\frac{\gamma_{en}-\gamma_{pn}}{\gamma_{cn}}\right)\times\vec{\nabla}\mu_c\nonumber\\&&
-{1\over ec}\vec{\nabla}\left({\gamma_{en}\mu_p-\gamma_{pn}\mu_e}\over{\gamma_{cn}}\right)\times\vec{\nabla}\psi,
\end{eqnarray} 
where we define the \emph{Hall drift velocity}, which is proportional to the electrical current density, by 
 
\begin{equation} 
\label{Hall} 
\vec{v}_H\equiv \frac{\gamma_{en}-\gamma_{pn}}{\gamma_{cn}}(\vec{v}_p-\vec{v}_e)=\frac{c(\gamma_{en}-\gamma_{pn})}{n_c e \gamma_{cn}}\vec{J}. 
\end{equation} 
The first term on the RHS of Eq.~(\ref{magn}) shows that the magnetic field is transported by the sum of the neutron velocity $\vec{v}_n$, the ambipolar diffusion velocity $\vec{v}_A$, and the Hall drift velocity $\vec{v}_H$. The second term represents the ohmic diffusion, where the electrical conductivity is  
 
\begin{equation} 
\sigma=e^2\left[\gamma_{ep}+\frac{n_n}{n_c}\left(\frac{1}{\gamma_{pn}}+\frac{1}{\gamma_{en}}\right)^{-1}\right]^{-1}. 
\end{equation} 
 
\noindent Finally, the last two terms on the RHS of Eq.~(\ref{magn}) represent battery effects.  
 
To complete a set of equations to describe the system evolution, we need equations for the evolution of the particle densities. Weak interactions between the particles tend to reduce chemical-potential imbalances between the charged particles and neutrons. We define the differences between the rates, per unit volume, of the reactions $p+e\rightarrow n+\nu_e$ and $n\rightarrow p+e+\overline{\nu_e}$ as $\Delta \Gamma \equiv\Gamma(p+e\rightarrow n+\nu_e)-\Gamma(n\rightarrow p+e+\overline{\nu_e})=\lambda \Delta \mu$, where $\Delta \mu= \mu_c-\mu_n$ is the chemical imbalance, and the parameter $\lambda$ depends generally on density and temperature (GR-92) \footnote{If $\Delta \mu \gtrsim 5kT$, where $k$ is the Boltzmann constant and $T$ is the temperature, $\lambda$ must also be allowed to depend on $\Delta \mu$ (Reisenegger \cite{Rei-95}; Fern\'andez \& Reisenegger \cite{F-05}).}. The \emph{chemical quasi-equilibrium} between the charged particles and neutrons is reached when $\Delta \mu=0$. The continuity equations for the particle densities are then given by 
 
\begin{equation} 
\label{densn} 
\frac{\partial n_i}{\partial t}+\vec{\nabla}\cdot(n_i\vec{v_i})=\pm\lambda(\Delta \mu). 
\end{equation} 
 
\noindent In Eq.~(\ref{densn}), the $+$ sign corresponds to the neutrons and the $-$ sign to the electrons and protons. Defining the \emph{baryon} number density as $n_B\equiv n_n+n_c$ and adding Eq.~(\ref{densn}) for charged particles and neutrons, we obtain the conservation law for $n_B$, 
 
\begin{equation} 
\label{densb} 
\frac{\partial n_B}{\partial t}+\vec{\nabla}\cdot(n_n\vec{v_n}+n_p\vec{v_p})=0. 
\end{equation}

\noindent The set of differential Eqs. (\ref{magn}), (\ref{densn}), and (\ref{densb}) are first, rather than second order in time, that is they involve the velocities of the three fluid species, but not their accelerations. 
 
\section{One-Dimensional Model} 
\label{1dmodel} 
 
\subsection{Basic Equations}
\label{bequ} 
 
We consider a one-dimensional geometry in which the magnetic field points in one Cartesian direction $z$, but varies only along an orthogonal direction $x$ as $\vec{B}(\vec{r},t)=B_z(x,t)\hat{z}$. We also assume that all of the other physical variables vary only along $x$, and therefore that the gradient operator is $\vec{\nabla}=\hat{x}(\partial/\partial x)$. From Ampere's law, $J_x=(c/4\pi)(\vec{\nabla}\times\vec{B})_x=0=n_ce(v_{px}-v_{ex})$, thus, $v_{ex}=v_{px}\equiv v_{cx}=v_{nx}+v_{Ax}$. Using Eqs.~ (\ref{ambi}), (\ref{magn}), (\ref{Hall}), (\ref{densn}), and (\ref{densb}), and defining $r\equiv c^2/(4\pi\sigma)$, we obtain the following non-linear set of equations for the evolution of the magnetic field and particle densities:  
 
\begin{equation} 
\label{1dmagne} 
\frac{\partial B_{z}}{\partial t}=-\frac{\partial (v_{cx}B_z)}{\partial x}+\frac{\partial }{\partial x}\left(r\frac{\partial B_z}{\partial x}\right), 
\end{equation} 
 
\begin{equation} 
\label{nB1d} 
\frac{\partial n_B}{\partial t}= -\frac{\partial}{\partial x}\left(n_n v_{nx}+n_c v_{cx}\right), 
\end{equation}

\begin{equation} 
\label{nc1d} 
\frac{\partial n_c}{\partial t}=-\frac{\partial}{\partial x}\left(n_cv_{cx}\right)-\lambda \left(\Delta \mu \right), 
\end{equation} 
 
\noindent where

\begin{eqnarray} 
v_{nx}=-\frac{1}{\alpha n_n}\left[n_{c} \left(\frac{\partial \mu_{c}}{\partial x}+\frac{\mu_{c}}{c^2}\frac{\partial \psi}{\partial x}\right)+n_{n}\left(\frac{\partial \mu_{n}}{\partial x}+\frac{\mu_{n}}{c^2}\frac{\partial \psi}{\partial x}\right)\right. \nonumber\\
\left.+\frac{\partial}{\partial x}\left(\frac{B_z^2}{8\pi}\right)\right],
\end{eqnarray} 
 
\noindent and 
 
\begin{equation} 
v_{Ax}=-\frac{1}{n_n n_c\gamma_{cn}}\left[n_{c} \left(\frac{\partial \mu_{c}}{\partial x}+\frac{\mu_{c}}{c^2}\frac{\partial \psi}{\partial x}\right)+\frac{\partial}{\partial x}\left.(\frac{B_z^2}{8\pi}\right)\right]. 
\end{equation} 
In this one-dimensional geometry $\vec{\nabla}\times \left(\vec{v}_H\times \vec{B}\right)=\vec{0}$, and, because we are considering only variations along $x$, the gradients that appear in the battery terms of Eq.~(\ref{magn}) are parallel. Thus, there is no contribution from the Hall drift and the battery terms to the evolution of the magnetic field.  
 
\subsubsection {Conserved quantities and boundary conditions}
\label{conserve} 
 
We note that Eqs.~(\ref{1dmagne}) and (\ref{nB1d}) can be written as flux-conserving equations of the form  
 
\begin{equation} 
\label{consb} 
\frac{\partial A}{\partial t}=-\frac{\partial S_A}{\partial x}, 
\end{equation} 
 
\noindent where $A$ represents either $B_z$ or $n_B$, and $S_A$ is defined in each case by 
 
\begin{equation} 
S_{B_z}=v_{cx}B_z-\frac{c^2}{4\pi\sigma}\frac{\partial B_z}{\partial x}, 
\end{equation} 
 
\noindent and 
 
\begin{equation} 
S_{n_B}=n_n v_{nx}+n_cv_{cx}. 
\end{equation} 
 
On the other hand, if we define $C_{A}=\int_0^{x_{max}}A(x,t)dx$, we obtain the conservation law

\begin{equation} 
\label{conserv} 
\frac{d}{dt}C_A=S_A|_{x=0}-S_A|_{x=x_{max}},  
\end{equation} 
 
\noindent We note that $C_{B_z}\equiv \Phi_B(t)$ is the magnetic flux, while $C_{n_B}\equiv N_B(t)$ is the baryon number. We see from Eq.~(\ref{conserv}) that $C_A$ is a conserved quantity if $S_A|_{x=0}-S_A|_{x=x_{max}}=0$, which depends on the boundary conditions imposed on the variables contained in $S_A$. To conserve both the magnetic flux and the baryon number during evolution, we impose the boundary conditions 
 
\begin{equation} 
v_{cx}(x=0,t)=v_{cx}(x_{max},t)=0, 
\end{equation} 
 
\begin{equation} 
v_{nx}(x=0,t)=v_{nx}(x_{max},t)=0, 
\end{equation} 
 
\begin{equation} 
\frac{\partial B_z}{\partial x}(x=0,t)=\frac{\partial B_z}{\partial x}(x=x_{max},t)=0. 
\end{equation} 
Of course, this restriction does not provide a realistic description of a neutron star, in which magnetic flux can be lost through the boundaries; however, it allows us to control the precision of the calculation in this one-dimensional case. Although this geometry is quite unrealistic, it enables us to study the timescales involved in an analytical and numerical fashion. In the future, when we study this system in higher numbers of dimensions, we will use more appropiate boundary conditions for the magnetic field.

\subsection{Linearization of the equations with respect to the densities}
\label{lineardens} 
 
Since in neutron star conditions, the ratio between the magnetic pressure $B_z^2/8\pi$ and the degeneracy pressure of the charged particles is very small, we consider the magnetic field to be a small perturbation on a non-magnetized background in full hydrostatic, chemical, and diffusive quasi-equilibrium. If we label the physical variables characterizing the background by the sub-index $0$, the hydrostatic, diffusive, and chemical equilibria are given by 
 
\begin{equation} 
\label{hydro1} 
\frac{\partial \mu_{0c}}{\partial x}+\frac{\mu_{0c}}{c^2}\frac{\partial \psi_0}{\partial x}=0, 
\end{equation} 
 
\begin{equation} 
\label{hydro2} 
\frac{\partial \mu_{0n}}{\partial x}+\frac{\mu_{0n}}{c^2}\frac{\partial \psi_0}{\partial x}=0, 
\end{equation} 
 
\begin{equation} 
\label{chemi} 
\Delta \mu_0=\mu_{0c}-\mu_{0n}=0, 
\end{equation} 
 
The magnetic field cannot force significant displacements of the particles; we therefore write $n_i(x,t)= n_{0i}(x)+\delta n_i(x,t)$, where $|\delta n_i(x,t)|\ll n_{0i}(x)$. The chemical potentials are $\mu_i(x,t)=\mu_{0i}(x)+\delta \mu_i(x,t)$, where $\delta \mu_i(x,t)=k_{iB}\delta n_B(x,t)+k_{ic}\delta n_c(x,t)$, $k_{iB}=\left(\partial \mu_{i}/\partial n_{B}\right)_{n_{0c}}$, and $k_{ic}=\left(\partial \mu_{i}/\partial n_{c}\right)_{n_{0B}}$, with $k_{eB}=0$ and the remaining coefficients are calculated from the equation of state. We also use the Cowling approximation, neglecting the perturbations of the gravitational potential with respect to the background value, i.e., $\psi=\psi_0$. By neglecting terms of second order in $\delta n_i$ and using that $v_{cx}=v_{nx}+v_{Ax}$, $ \delta n_{B}=\delta n_{n}+\delta n_{c}$, and the definitions $k_{0c}\equiv k_{cc}-k_{nc}$, $k_{cc}\equiv k_{pc}+k_{ec}$, $k_{0B}\equiv k_{pB}-k_{n_B}$, and $r_0 \equiv c^2/(4\pi \sigma_0)$, we derive the following set of equations, which are nonlinear with respect to the magnetic field $B_z(x,t)$, but linear with respect to the density perturbations $\delta n_B(x,t)$ and $\delta n_c(x,t)$, namely, 
 
\begin{equation} 
\label{bz} 
\frac{\partial B_z}{\partial t}=-\frac{\partial (v_{cx}B_z)}{\partial x}+\frac{\partial }{\partial x}\left(r_0 \frac{\partial B_z}{\partial x}\right), 
\end{equation} 
 
\begin{equation} 
\label{nB} 
\frac{\partial \delta n_B}{\partial t}= -\frac{\partial}{\partial x}\left(n_{0n}v_{nx}+n_{0c}v_{cx}\right), 
\end{equation}

\begin{equation} 
\label{nc} 
\frac{\partial \delta n_c}{\partial t}=-\frac{\partial}{\partial x}\left(n_{0c}v_{cx}\right)-\lambda \left( k_{0c} \delta n_c + k_{0B} \delta n_B\right), 
\end{equation} 
where

\begin{eqnarray} 
\label{vn} 
v_{nx}&=&-\frac{1}{\alpha n_{0n}}\left[n_{0n}\mu_{0n}\frac{\partial}{\partial x}\left(\frac{k_{n_B} \delta n_B+k_{nc} \delta n_c}{\mu_{0n}}\right)\right.\nonumber\\&&
\left.+n_{0c}\mu_{0n}\frac{\partial}{\partial x}\left(\frac{k_{cc}\delta n_c+k_{pB}\delta n_B}{\mu_{0n}}\right)+\frac{\partial}{\partial x}\left(\frac{B_z^2}{8\pi}\right)\right], 
\end{eqnarray} 
and 
 
\begin{eqnarray} 
\label{vA} 
v_{Ax}=-\frac{1}{n_{0n}n_{0c}\gamma_{cn}}\left[n_{0c}\mu_{0n}\frac{\partial}{\partial x}\left(\frac{k_{cc}\delta n_c+k_{pB}\delta n_B}{\mu_{0n}}\right)\right.\nonumber\\
\left.+\frac{\partial}{\partial x}\left(\frac{B_z^2}{8\pi}\right)\right]. 
\end{eqnarray} 
 
\subsection{Dimensionless equations} 
\label{dimensionequ} 

We proceed in writing the set of Eqs. (\ref{bz}), (\ref{nB}), (\ref{nc}), (\ref{vn}), and (\ref{vA}) in terms of dimensionless variables. We start by defining dimensionless variables as $\overline{a}=a/a_s$, with the sub-index $s$ that represents a characteristic value of the corresponding dimensional variable. In the following, we explain the meaning of each of the characteristic values. We normalize the space variable $x$ with respect to the length of the system, that is the length of our computational domain $d$, thus $x_s=d$. We note, however, that there is another characteristic length $L$, which is the length over which the functions vary spatially; we therefore use this scale when dealing with order-of-magnitude estimates of spatial derivatives. The time variable $t$ is normalized with respect to $t_s\equiv \alpha d^2/(n_{0n}k_{nB})$. The meaning of this quantity will become clearer in the following section; it is related to the shortest relevant timescale of the system. We assume that $B_s=B_{z}^{max}$ is the maximum value of the initial magnetic field, $n_s=n_{0n}^{max}$ is the maximum neutron background density, $n_{cs}=n_{0c}^{max}$ is the maximum of the charged background density, $k_s=k_{nB}^{max}$ and $\mu_s=\mu_{0n}^{max}$ are properties of the background, $r_s=r_0^{max}$ is the maximum resistivity in the background. From the third term on the RHS of Eq.~(\ref{vn}, we estimate a characteristic order-of-magnitude value for the neutron velocity induced by the magnetic stress to be $v_s=B_s^2/(8\pi n_s x_s \alpha)$. If we compare the first and third parentheses on the RHS of Eq.~(\ref{vn}), and use that close to magnetohydrostatic quasi-equilibrium $\delta n_B/n_{0B}\sim \delta n_c/n_{0c}$, $n_{0n}/n_{0B}\sim 1$, and $n_{0n}k_{nB}\gg n_{0c}k_{n_c}$, then we can estimate a characteristic neutron-density perturbation to be $\delta n_s=B_s^2/(8\pi n_s k_s)$. Finally, defining $\overline{n}_{cs}=n_{cs}/n_{s}$ and $\overline{k}=k/k_s$ where $k$ is any of the corresponding parameters defined in the last section, we obtain the following dimensionless set of equations, 
 
\begin{equation} 
\label{magbar} 
\frac{\partial \overline{B}_z}{\partial \overline{t}}=-\overline{n}_{cs} \overline{\Upsilon}\frac{\partial (\overline{v}_{cx}\overline{B}_z)}{\partial \overline{x}}+\overline{\omega}\frac{\partial }{\partial \overline{x}}\left(\overline{r}_0 \frac{\partial \overline{B}_z}{\partial \overline{x}}\right), 
\end{equation}

\begin{equation} 
\label{nbnorm} 
\frac{\partial \delta \overline{n}_{B}}{\partial \overline{t}}= -\frac{\partial}{\partial \overline{x}}\left(\overline{n}_{0n} \overline{v}_{nx}+ \overline{n}_{0c}\overline{v}_{cx}\right), 
\end{equation}

\begin{equation} 
\label{ncnorm} 
\frac{\partial \delta \overline{n}_{c}}{\partial \overline{t}}=-\frac{\partial}{\partial \overline{x}}\left(\overline{n}_{0c}\overline{v}_{cx}\right)-\overline{\lambda}\overline{n}_{cs}\overline{\theta}\left(\overline{k}_{0c} \delta \overline{n}_{c}+\overline{k}_{0B} \delta \overline{n}_{B}\right), 
\end{equation} 
where 
 
\begin{eqnarray} 
\label{vnbar} 
\overline{v}_{nx}&=&-\frac{1}{\overline{n}_{0n}}\left[\overline{n}_{0n}\overline{\mu}_{0n}\frac{\partial}{\partial \overline x}\left(\frac{\overline{k}_{nB}\delta \overline{n}_B+\overline{k}_{nc}\delta \overline{n}_c}{\overline{\mu}_{0n}}\right)\right.\nonumber\\&&
\left.+\overline{n}_{0c}\overline{\mu}_{0n}\frac{\partial}{\partial \overline x}\left(\frac{\overline{k}_{cc}\delta \overline{n}_c+\overline{k}_{pB}\delta \overline{n}_B}{\overline{\mu}_{0n}}\right)+\frac{\partial(\overline{B}_z^2)}{\partial \overline{x}}\right], 
\end{eqnarray} 
 
\begin{eqnarray} 
\label{vabar} 
\overline{v}_{Ax}=-\frac{\overline{\epsilon}}{\overline{n}_{0n}\overline{n}_{0c}\overline{\gamma}_{cn}}\left[\overline{n}_{0c}\overline{\mu}_{0n}\frac{\partial}{\partial \overline x}\left(\frac{\overline{k}_{cc}\delta \overline{n}_c+\overline{k}_{pB}\delta \overline{n}_B}{\overline{\mu}_{0n}}\right)\right.\nonumber\\
\left.+\frac{\partial(\overline{B}_z^2)}{\partial \overline{x}}\right]. 
\end{eqnarray} 
In the above set of equations, the parameter $\overline \Upsilon \equiv B_s^2/(8\pi n_{0c_s} n_{0ns} k_{nB_s})$ controls the coupling strength between the magnetic field and the particle species, $\overline{\omega}\equiv \alpha r_{s}/(n_{s}k_{s})$ controls the importance of the ohmic diffusion, $\overline{\theta}\equiv \alpha \lambda_s x_s^2/n_{cs}$ controls the weak interaction rate, and finally $\overline{\epsilon} \equiv \alpha/(n_{s}\gamma_{s})$ controls the importance of the ambipolar diffusion process. In diffusive quasi-equilibrium, Eq.~(\ref{vn}) with $v_{nx}\approx 0$, and  Eq.~(\ref{vA}) with  $v_{Ax}\approx 0$, imply that 
 
\begin{equation} 
\label{fluctuch} 
\frac{\delta n_c}{n_{0c}} \sim \frac{B_z^2}{8\pi n_{0c}^2 (k_{cc}+k_{pB})}\sim \frac{B_s^2}{8\pi n_{0c_s}n_{0n_s}k_{nB_s}}\equiv\overline \Upsilon, 
\end{equation} 
where we used that the fraction $|n_{0n}k_{nB}|/|n_{0c}(k_{cc}+k_{pB})|\sim 1$. For instance, in the non-interacting particle limit of GR-92 this fraction is approximately equal to $2$ since the chemical potential of each particle species $i$ is a function only of its number density $\mu_i(\vec{r},t)=\mu_i(n_i(\vec{r},t))=\sqrt{(m_ic^2)^2+\hbar^2c^2[3\pi^2n_i(\vec{r},t)]^{2/3}}$, which implies that $k_{nB}=\left(\partial \mu_{n}/\partial n_{n}\right)_{n_{0n}}$, $k_{nc}=-k_{nB}$, $k_{pB}=0$, and $k_{ic}=\left(\partial \mu_{i}/\partial n_{c}\right)_{n_{0c}}$. On the other hand, for the background numerical values that we use in this paper this fraction is $\approx 0.7$.  These background parameters are extracted from the equation of state provided by Akmal et al. (\cite{A-98}), evaluated at a neutron number density $n_{0n}=9.8 \times 10^{37} cm^{-3},$ which corresponds to a charged-particle number density of $n_{0c}=3.9 \times 10^{36} cm^{-3}$ and to a total mass density (including neutrons, protons and electrons) of $1.5\times 10^{14}~g/cm^{3}.$ For these values,  $k_{nB}\approx 4.2\times 10^{-43} erg ~ cm^{3}$, $k_{nc}\approx -1.2\times 10^{-42} erg ~ cm^{3}$, $k_{pB}\approx -8.2\times 10^{-43} erg ~ cm^{3}$, $k_{pc}\approx 2.8\times 10^{-42} erg ~ cm^{3}$, and $k_{ec}\approx 1.3\times 10^{-41} erg ~ cm^{3}$. In addition, the linearization with respect to the variable $\delta n_c$ requires that $|\delta n_c|\ll n_{0c}$, i.e., $\overline{\Upsilon} \ll 1$ and therefore $B_s\ll 6.3\times 10^{16} G.$  
 
\section{Results} 
\label{res} 
 
\subsection{Characteristic evolutionary timescales} 
\label{carsca}
We consider a dynamical system of three independent variables controlled by three differential equations that are first order in time. Thus, in the linear limit, the system will have three exponentially decaying ``modes'' of different timescales. Although, this is not strictly true in the general nonlinear case, one can always identify three characteristic timescales on which the system approaches successive quasi-equilibrium states. Real neutron stars approach a magnetohydrostatic quasi-equilibrium on a timescale not much longer than the Alfv\'en time, which is of the order of seconds. Here, as described in Sect.~\ref{phys}, this early evolution is mimicked by a artificial friction force term proportional to a parameter $\alpha$; this parameter is chosen so that this timescale is sufficiently long enough for the numerical code to be able to deal with it, but shorter than the timescales of modeled processes, such as ambipolar diffusion and weak interactions. In what follows, we make analytic estimates of the characteristic evolutionary timescales and evaluate them for typical magnetar core parameters (see e.g., Arras et al. \cite{ACT-04}). In doing this, we can make sense of our numerical results and estimate the order of magnitude of the timescales involved in a real system, under the limitations of our one-dimensional model. 
  
\subsubsection{Timescale to achieve magnetohydrostatic quasi-equilibrium} 
\label{tmhdcar}
 
We assume that our non-magnetized background star is in hydrostatic quasi-equilibrium [see Eqs.~(\ref{hydro1}) and (\ref{hydro2})], in other words, the net force on a fluid element containing all species is zero. When a magnetic field is present in the system, the magnetic force applies pressure to the charged particles (electrons and protons), inducing density perturbations that create an imbalance between the different forces acting on a fluid element [see Eq.~(\ref{vn})]. The particles must move until a magnetohydrodystatic (MHS) quasi-equilibrium is reached. The neutron velocity necessary to achieve this balance is expressed in Eq.~(\ref{vn}). During the first stages of evolution, the collisional coupling between the charged particles and neutrons compels them move with about the same velocity, $v_{cx}\approx v_{nx}$. Neglecting weak interactions, from Eqs.~\ref{nbnorm} and \ref{ncnorm} we obtain the consistency condition 
 
\begin{equation} 
\label{mhdc} 
\delta n_B / n_{0B} \approx \delta n_c/n_{0c} \ll 1.  
\end{equation} 
The induced charged-particle and neutron-pressure gradients tend to choke the magnetic force. The magnetohydrostatic quasi-equilibrium state is reached when there is a close balance between these opposing forces [see RHS of Eq.~(\ref{vn})], that is, 
  
\begin{equation} 
\label{condi1} 
n_{0n}\left(k_{nB}+\frac{n_{0c}}{n_{0B}}k_{nc}\right)\delta n_B \sim \frac{B_z^2}{8\pi}, 
\end{equation}  
where we neglected the second term in the RHS of Eq.~(\ref{vn}) since $n_{0c}\ll n_{0n}$ and used the Eq.~(\ref{mhdc}). The velocity induced by the initially-unbalanced magnetic-pressure gradient is  
 
\begin{equation} 
\label{condi2} 
v_{nx}\approx v_{cx}\sim \frac{B_z^2}{8\pi\alpha n_{0n}L}.  
\end{equation} 
On the other hand, from Eq.~(\ref{nB}), we obtain the time required to create (or destroy) a perturbation $\delta n_B$ as 
\begin{equation} 
\label{nn} 
t_{MHS}\sim\frac{L\delta n_B}{n_{0B}v_{nx}}. 
\end{equation} 
Using Eqs.~(\ref{condi1}), (\ref{condi2}), and (\ref{nn}), and $n_{0B}\sim n_{0n}$ we estimate the timescale to reach the magnetohydrostatic quasi-equilibrum as 
 
\begin{equation} 
\label{tnn} 
t_{MHS}\sim \frac{\alpha L^2}{n_{0n}k_{nB}\left(1+\frac{n_{0c}}{n_{0n}}\frac{k_{nc}}{k_{nB}}\right)}\approx \frac{\alpha L^2}{n_{0n}k_{nB}}=\left(\frac{L}{d}\right)^2 t_s, 
\end{equation} 
thus, $\overline{t}_{MHS}\sim (L/d)^2 \leq 1$. In Eq.~(\ref{tnn}) we used that $|(n_{0c}k_{nc})/(n_{0n}k_{nB})|\ll 1$. In fact, in the non-interacting particle limit of GR-92 we have $k_{nc}=-k_{nB}$, thus, this fraction is of the order of $|n_{0c}/n_{0n}|\ll 1$. For the numerical background values that we use in this paper, this fraction is $\approx 0.12$. We note that the scale $t_{MHS}$ is the shortest relevant timescale in the system, and is controlled by the artificial $\alpha$ parameter that was introduced in our slow-motion approximation (Sect.~\ref{phys}). Real neutron stars evolve to the magnetohydrostatic quasi-equilibrium on a short timescale not much longer than the Alfv\'en time, which for typical magnetar core parameters scales as    
 
\begin{equation} 
\label{taf} 
t_{Alfven}=5.7~\times 10^{-2}~ R_{6}~B_{15}^{-1}~s, 
\end{equation}   
where $R_{6}\equiv R/(10^6~cm)$ denotes the radius of the star in units of $10^6~cm,$ and $B_{15}\equiv B_z/(10^{15}~G)$ the magnetic field in units of $10^{15}~G$ (this timescale as well as the following ones are evaluated at the typical mass density $1.5\times 10^{14}~g/cm^{3}$). 
This timescale is far shorter than the timescales of the processes that promote the long-term evolution of the magnetic field (see Sects.~\ref{tcardiff}-\ref{tmagnf}), which are of the order of years or much longer. A numerical code simulating the evolution on the Alfv\'en timescale would require a time step many orders of magnitude shorter than that required to simulate the long-term evolution in a computational time that is not prohibitively long. We overcome this difficulty by replacing the short-term dynamics by the artificial friction term proportional to the parameter $\alpha$. This parameter is chosen so that the timescale in Eq.~(\ref{tnn}) is long enough for the numerical code to be able to deal with it (and therefore much longer than the Alfv\'en time), but shorter than the timescales of the long-term processes that we discuss in the following sections.

\subsubsection{Timescale for charged particles to reach diffusive quasi-equilibrium through ambipolar diffusion neglecting weak interactions} 
\label{tcardiff}
 
We now assume that the magnetohydrostatic quasi-equilibrium, discussed in the last section, has been reached. However, the charged particles continue to move relative to the neutrons due to ambipolar diffusion (subject to collisional drag among different species), and they reach diffusive quasi-equilibrium when there is a close balance between the magnetic force and the charged-particle pressure gradients (see RHS of Eq.~(\ref{vA}). By an argument analogous to that in the previous section and using the temperature dependence of the collisional frequencies from Ref. \cite{YS-90}, we obtain the timescale $t_{drag}$ for charged particles to reach diffusive quasi-equilibrium as 
 
\begin{equation} 
\label{tncc} 
t_{drag}\sim \frac{n_{0n}\gamma_{cn}L^2}{n_{0c}(k_{cc}+k_{pB})}\sim \frac{\gamma_{cn}L^2}{k_{nB}}\sim~4.5~\times 10^{-1}~ L_5^2~T_8^2 ~ yr, 
\end{equation} 
where $L_5\equiv L/(10^5~cm)$ and $T_8\equiv T/(10^8 K).$ We can write this time in units of our normalization time $t_s$ as 
 
\begin{equation}
\label{bartnc} 
\overline{t}_{drag}\sim\left(\frac{L}{d}\right)^2\frac{n_{0n}\gamma_{cn}}{\alpha} \sim\frac{\overline{t}_{MHS}}{\overline {\epsilon}}\sim\left(\frac{L}{d}\right)^2\frac{1}{\overline{\epsilon}},
\end{equation} 
We require $\overline{t}_{drag}\gg \overline{t}_{MHS}$, therefore $\overline \epsilon \ll 1 $. The $\overline \epsilon$ parameter in Eq.~(\ref{bartnc}), which is inversely proportional to the collisional parameter $\gamma_{cn}$, controls the timescale on which the charged particles reach the diffusive quasi-equilibrium. 
 
\subsubsection{Timescale to achieve chemical quasi-equilibrium through weak interactions, neglecting ambipolar diffusion} 
\label{tchemcar} 
If there is a perturbation of the chemical quasi-equilibrium ($k_{0c}\delta n_c \neq -k_{0B}\delta n_B$; see RHS of Eq.~(\ref{nc}), the characteristic timescale on which the chemical quasi-equilibrium is restored through weak interactions (charged particles decaying into neutrons and viceversa) can be estimated from Eq.~(\ref{nc}). Neglecting the first term on the RHS, which takes into account the ambipolar diffusion, comparing the terms with $\delta n_c$, and using the temperature dependence of the $\lambda$ parameter from Ref.\cite{S-89}, which assumes modified Urca reactions (e.g., Ref. \cite{S-83}), we obtain 
 
\begin{equation} 
\label{dmu1} 
t_{weak}\sim \frac{1}{\lambda (k_{0c}+k_{0B})}\sim 4.3 ~ \times 10^5~ T_8^{-6} ~yr. 
\end{equation} 
 
If we write Eq.~(\ref{dmu1}) in units of our normalization time, we derive 
 
\begin{equation} 
\label{bartdmu} 
\overline{t}_{weak}\sim \frac{n_{0n}k_{nB}}{\alpha \lambda d^2 (k_{0c}+k_{0B})}\sim \frac{n_{0c}}{\alpha \lambda d^2}\sim \frac{1}{\overline{\theta}}. 
\end{equation} 
Using Eq.~(\ref{bartdmu}), we obtain $\overline{t}_{weak}/\overline{t}_{MHS}\sim (d/L)^2 (1/\overline{\theta})$; we therefore require $\overline \theta \ll 1$, so that $\overline{t}_{weak}\gg \overline{t}_{MHS}$. The $\overline \theta$ parameter in Eq.~(\ref{bartdmu}) is directly proportional to the weak interaction rate parameter $\lambda$, which controls the timescale on which the chemical quasi-equilibrium is achieved.  
 
Both weak interactions and ambipolar diffusion processes contribute in general to the decay of the charged-particle density perturbations. The more rapidly-acting processes determine the evolutionary timescale of the charged particles. We propose an approximate general interpolation formula: 
 
\begin{equation} 
\label{dmu} 
\overline t_{\delta n_c}\sim \left[\frac{1}{\overline{t}_{drag}}+ \frac{1}{\overline{t}_{weak}}\right]^{-1}\sim  \left[\left(\frac{d}{L}\right)^2 \overline {\epsilon}+ \overline{\theta}\right]^{-1}. 
\end{equation} 
 
\subsubsection{Ohmic diffusion timescale} 
\label{tohmcar}  
The timescale on which the magnetic field decays by ohmic diffusion can be estimated from Eq.~(\ref{bz}), neglecting the first term on the RHS and using the temperature dependence of the electrical conductivity from Ref. \cite{HUY-90}, as  
 
\begin{equation} 
\label{tresis} 
t_{ohmic}\sim \frac{L^2}{r_0}=\frac{4\pi\sigma_0 L^2}{c^2}~\sim 1.4~\times~ 10^{11}~ L_5^{2}~ T_8^{-2}~yr. 
\end{equation} 
 
On the other hand, using Eq.~(\ref{tnn}), we obtain 
 
\begin{equation} 
\label{bartohm} 
\overline{t}_{ohmic}\sim \left(\frac{L}{d}\right)^2 \frac{n_{0n}k_{nB}}{\alpha r_0} \sim \frac{\overline{t}_{MHS}}{\overline{\omega}}\sim \left(\frac{L}{d}\right)^2\frac{1}{\overline{\omega}}. 
\end{equation} 
The $\overline \omega$ parameter is directly proportional to the parameter $r_0$ and controls the ohmic diffusion timescale. We require $\overline \omega \ll 1 $, so that $\overline{t}_{ohmic}\gg \overline{t}_{MHS}$.

\subsubsection{Timescale for the evolution of the magnetic field by ambipolar diffusion with weak interactions} 
\label{tmagnf} 
The two terms on the RHS of Eq.~(\ref{magbar}) set the two basic timescales on which the magnetic field decays. The first of these terms couples the magnetic-field evolution with the particle dynamics, while the second one gives the magnetic-field evolution due to ohmic diffusion. For the remainder of this section, we estimate the magnetic-field evolution timescale by neglecting the second term (formally we set $\overline \omega=0$ or $\overline{t}_{ohmic}\rightarrow \infty$). Again, it is useful to distinguish between the two opposite extreme regimes discussed in Sects.~\ref{tcardiff} and ~\ref{tchemcar}. 
 
In the first regime, the ambipolar diffusion occurs more rapidly than the weak-interaction processes, that is $t_{MHS}  \ll t_{drag} \ll t_{weak}$. First, the system reaches the magnetohydrostatic quasi-equilibrium in the short timescale $t_{MHS}$. In the second stage, all particles reach the diffusive quasi-equilibrium in the timescale $t_{drag}$  and there is a close balance between the Lorentz force and the charged particle pressure gradient [see Eq.~(\ref{vA})]. Since the magnetic field generates density perturbations in the charged particles but not in the neutrons, it prevents the system from reaching chemical quasi-equilibrium. Weak interactions tend to restore the local chemical quasi-equilibrium on a characteristic timescale $t_{weak}$. This tendency causes a slight reduction of the pressure gradient in the charged particles with respect to the Lorentz force, compelling the charged particles and the magnetic flux to move together at a small ambipolar velocity $v_{Ax}\approx v_{cx}$. This movement maintains the charged-particle pressure gradients, which tend to be erased by weak interactions that continue to operate. This interplay of weak interactions and ambipolar diffusion stops only once the pressure and magnetic field gradients disappear. 
 
To estimate the timescale of this process, we first note that the ambipolar velocity needed to keep the charged-particle density perturbations stationary with respect to time ($\partial \delta \overline{n}_{c}/\partial \overline{t}=0$) can be estimated from Eqs.~(\ref{nc}) and (\ref{fluctuch}) as 
 
\begin{equation} 
\label{vaequi} 
v_{Ax}\sim \frac{LB_z^2 \lambda (k_{0c}+k_{0B})}{8\pi n_{0c}n_{0n}k_{nB}}.  
\end{equation}  
Since we neglect ohmic diffusion, the magnetic-field evolution is governed by the coupling with the charged-particle movement at the ambipolar velocity given by  Eq.~(\ref{vaequi}). From Eq.~(\ref{bz}), and using Eq.~(\ref{vaequi}), we can estimate the timescale on which the magnetic field evolves in this case to be 
 
\begin{eqnarray} 
\label{tb} 
t_{ambip}^{(1)}\sim \frac{L}{v_{Ax}} \sim  \frac{{t}_{weak}}{\overline \Upsilon} \sim \frac{8\pi n_{0c}n_{0n}k_{nB}}{B_z^2 \lambda (k_{0c}+k_{0B})}\nonumber\\
~\sim~ 1.7\times~ 10^{9}~ B_{15}^{-2}~ T_8^{-6}~yr. 
\end{eqnarray}  
 
Using Eqs.~(\ref{fluctuch}) and (\ref{tb}), we write this timescale in dimensionless form as 
 
\begin{equation} 
\label{tb2} 
\overline{t}_{ambip}^{(1)}\sim \frac{1}{\overline{\Upsilon}~\overline{\theta}}. 
\end{equation}  
 
In this regime, the weak interactions operate very slowly, converting particles of one species into the other (beta decays) in a tendency to erase the chemical imbalance. The beta decays perturb the diffusive quasi-equilibrium, promoting magnetic-field evolution by means of ambipolar diffusion in a timescale that is limited by how rapidly the weak interactions operate ($t_{weak}$). We note that the timescale given by Eq.~(\ref{tb}) is the timescale to reach chemical quasi-equilibrium, $t_{weak},$ amplified by the factor $1/\overline{\Upsilon}$ (which is of the order of the ratio of the charged-fluid pressure to the magnetic pressure). The dependence of the timescale on this factor is expected since the Lorentz force drives the ambipolar diffusion, which maintains the chemical imbalance as long as there is a magnetic-field gradient. 
 
In the opposite regime, the weak interactions occur much faster than the ambipolar diffusion process, i.e. $t_{MHS} \ll t_{weak} \ll t_{drag}$. As before, during the first stage of this regime, the system reaches the magnetohydrostatic quasi-equilibrium in the short timescale $t_{MHS}$. At the end of this stage, the gradient of the magnetic pressure is balanced by the combined degeneracy pressure of all particles, with the neutrons providing the main contribution due to the much higher density. During the second stage, chemical quasi-equilibrium is established at every point in the system, coupling the neutron and charged-particle density perturbations. Since the charged-particle pressure is far smaller than that of the neutrons, it can not by itself balance the magnetic-pressure gradient, which, according to Eq.~(\ref{vA}), causes an ambipolar drift of velocity 
 
\begin{equation} 
\label{vaaprox} 
v_{Ax}\sim \frac{B_z^2}{n_{0n}n_{0c}\gamma_{cn}8\pi L}. 
\end{equation}  
 
On the other hand, from Eq.~(\ref{bz}) we have $t_{ambip}\sim L/v_{Ax}$, thus, we can estimate the magnetic-field evolutionary timescale in this case to be 
 
\begin{eqnarray} 
\label{tambi22} 
t_{ambip}^{(2)}\sim \frac{L}{v_{Ax}}\sim \frac{t_{drag}}{\overline \Upsilon}\sim \frac{n_{0n}n_{0c}\gamma_{cn}8\pi L^2}{B_z^2}\nonumber\\
~\sim 1.8\times 10^{3}~ B_{15}^{-2}~ L_{5}^2 ~ T_8^{2}~yr. 
\end{eqnarray}

If we use Eq.~(\ref{bartnc}), we obtain 
 
\begin{equation} 
\label{tambapr} 
\overline{t}_{ambip}^{(2)}\sim \left(\frac{L}{d}\right)^2\frac{1}{\overline{\Upsilon}~\overline{\epsilon}}. 
\end{equation} 
 
In this case, the beta decays restore the chemical quasi-equilibrium quickly but the collisions between particles are frequent. This prevents the particles from moving easily relative to each other to achieve diffusive quasi-equilibrium. In this regime, the main mechanism promoting magnetic-field evolution is ambipolar diffusion. This process carries the magnetic field with a velocity that is limited by the collision rate between particles and depends on the magnetic-field strength. Equation.~(\ref{tambi22}) gives the magnetic-field evolutionary timescale in this regime, which is that required for the particles to reach diffusive quasi-equilibrium, $t_{drag},$ but, as above, amplified by the factor $1/\overline{\Upsilon}$.

We conclude that, in general, where both weak interations and ambipolar diffusion contribute to the magnetic-field evolution, the slower of these processes determines the evolutionary timescale of the magnetic field. For a general estimate, we use the approximate interpolation formula that recovers the timescales of the regimes discussed above to the corresponding limits 
 
\begin{equation} 
\label{GRe92} 
t_{ambip}\sim t_{ambip}^{(1)}+t_{ambip}^{(2)}. 
\end{equation} 
Equation~(\ref{GRe92}) corresponds to the estimate of  Eq.~(35) in GR-92 in the limit of non-interacting particles. For a case that includes ohmic diffusion, the shorter of the timescales given in, Eqs.~(\ref{tresis}) and (\ref{GRe92}) provides the timescale for magnetic-field evolution, which we estimate using the interpolation formula

\begin{equation} 
\label{tbch1} 
t_B\sim \left[\frac{1}{t_{ambip}}+ \frac{1}{t_{ohmic}}\right]^{-1}, 
\end{equation} 
that can be written in dimensionless form as

\begin{equation} 
\label{tbch} 
\overline{t}_B\sim \left[\left\{\frac{1}{\overline \Upsilon}\left[\left(\frac{L}{d}\right)^2\frac{1}{\overline \epsilon}+\frac{1}{\overline \theta} \right]\right\}^{-1}+ \left(\frac{d}{L}\right)^2\overline \omega \right]^{-1}. 
\end{equation} 
 
The ordering between $\overline{t}_{\delta n_c}$ and $\overline{t}_B$ depends on the relative size of $\overline \omega$ and the other dimensionless parameters. We can place the relevant timescales in increasing order as 
 
\begin{equation} 
\label{torder1} 
\overline{t}_{1}=\overline{t}_{MHS}, 
\end{equation}

\begin{equation} 
\label{torder2} 
\overline{t}_{2}=\min\{\overline{t}_{\delta n_c},\overline{t}_{B}\}\approx \left(\frac{1}{\overline{t}_{\delta n_c}}+\frac{1}{\overline{t}_{B}}\right)^{-1}, 
\end{equation}

\begin{equation} 
\label{torder3} 
\overline{t}_{3}=\max\{\overline{t}_{\delta n_c},\overline{t}_{B}\}\approx \left(\overline{t}_{\delta n_c}+\overline{t}_{B}\right), 
\end{equation} 
where $\overline{t}_{1}\ll \overline{t}_{2} < \overline{t}_{3}$. It is easy to show that, for $\overline \omega=0$, the ordering between the timescales is $\overline{t}_{1}=\overline{t}_{MHS}$, $\overline{t}_{2}=\overline{t}_{\delta n_c}$, and  $\overline{t}_{3}=\overline{t}_{B}$.

\subsection{Normal Modes}
\label{normmodes} 
 
The set of Eqs.~(\ref{magbar}), (\ref{nbnorm}), and (\ref{ncnorm}) is nonlinear with respect to the magnetic field variable $B_z$. To find a linear solution of this set, we assume that the properties of the background star are homogeneous, i.e., quantities with sub-index zero do not depend on position, and we linearize the magnetic field as 
 
\begin{equation} 
\label{linearb} 
B_z(x,t)=B_{c}+\delta B_z(x,t),  
\end{equation} 
 
\noindent where $B_{c}$ is a constant and $|\delta B_z(x,t)| \ll B_c$ is a small magnetic perturbation. If we choose $B_c$ as the magnetic-field normalization unit (i.e., $\overline B_z=B_z/B_c$), we obtain 
 
\begin{equation} 
\label{bpert1} 
\overline{B}_z(\overline x, \overline t)=1+\delta \overline{B}_z(\overline x, \overline t). 
\end{equation} 
 
\begin{figure} 
\centering 
\resizebox{9cm}{9cm}{\includegraphics{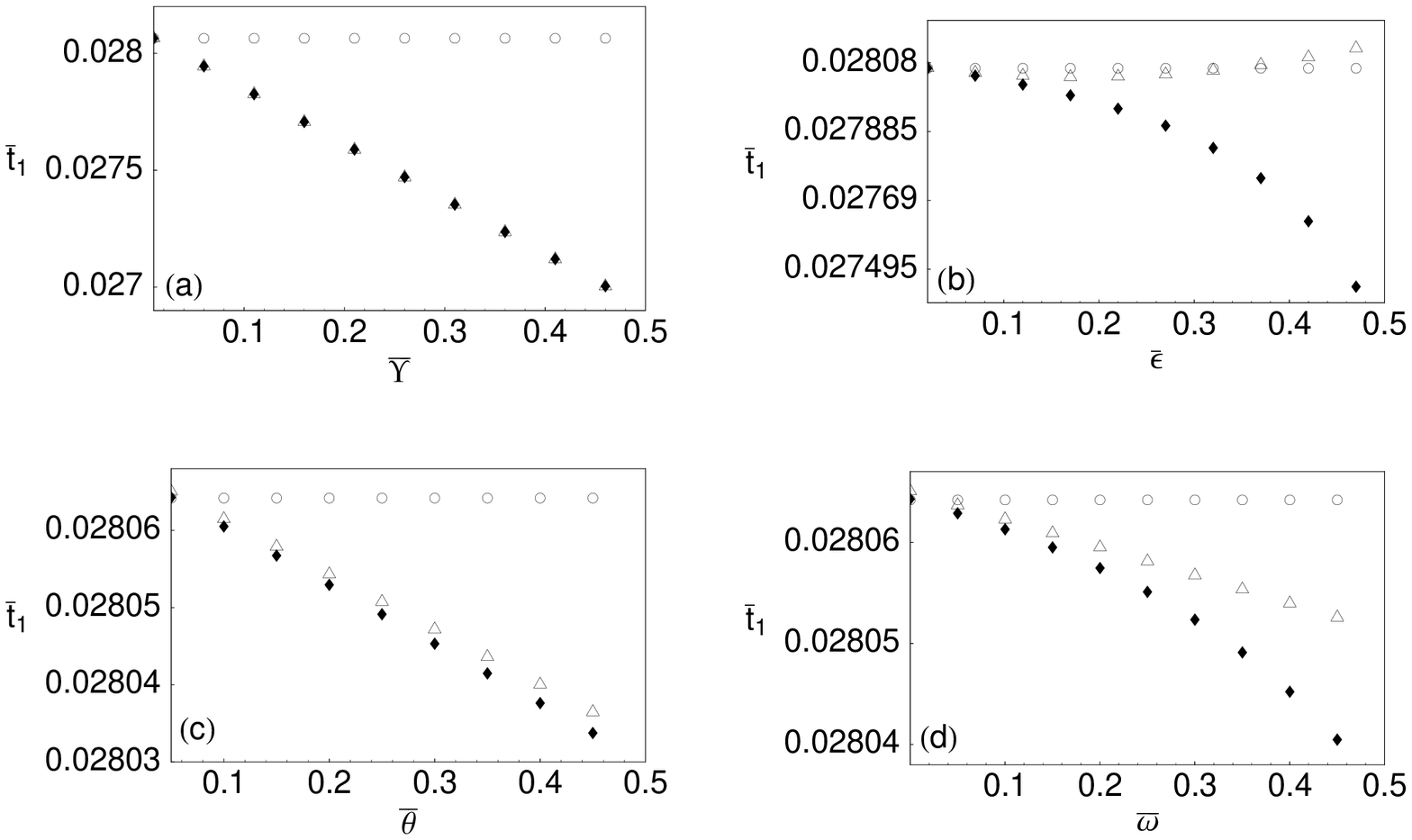}} 
\caption{Characteristic time $\overline{t}_1$ on which the system achieves magnetohydrostatic quasi-equilibrium, as a function of the dimensionless control parameters of the different physical processes (i.e., $\overline \Upsilon$~ [magnetic field strength], $\overline \epsilon$~ [collision rate], $\overline \theta$~ [weak interaction rate] and $\overline \omega$~ [ohmic resistivity]). The time variable $t$ is normalized as $\overline t = t/t_s$ where $t_s=\alpha d^2/(n_{0n}k_{nB})$ and $\alpha$ is an artificial parameter controlling how rapidly the system reaches magnetohydrostatic quasi-equilibrium and $k_{nB}$ is a property of the background system. We show a comparison of the characteristic time obtained from the order-of-magnitude estimate of Eq.~(\ref{torder1lin}) (open circles) with those obtained from the numerical solution of the eigenvalue problem of 
Eq.~(\ref{eigen}) (black diamonds) and from the analytical approximation of Eq. ~(\ref{val1}) (open triangles). (a)  $\overline{t}_1$ as a function of $\overline \Upsilon$ for $\overline \epsilon=0.02$, $\overline \theta=0.05$, and $\overline \omega=0.0001$. (b)  $\overline{t}_1$ as a function of $\overline \epsilon$ for $\overline \Upsilon=0.01$, $\overline \theta=0.05$, and $\overline \omega=0.0001$. (c)  $\overline{t}_1$ as a function of $\overline \theta$ for $\overline \epsilon=0.02$, $\overline \Upsilon=0.01$, and $\overline \omega=0.0001$. (d) $\overline{t}_1$  as a function of $\overline \omega$ for $\overline \epsilon=0.02$, $\overline \Upsilon=0.01$, and $\overline \theta=0.05$.} 
\label{t1} 
\end{figure}

\begin{figure} 
\centering 
\resizebox{9cm}{9cm}{\includegraphics{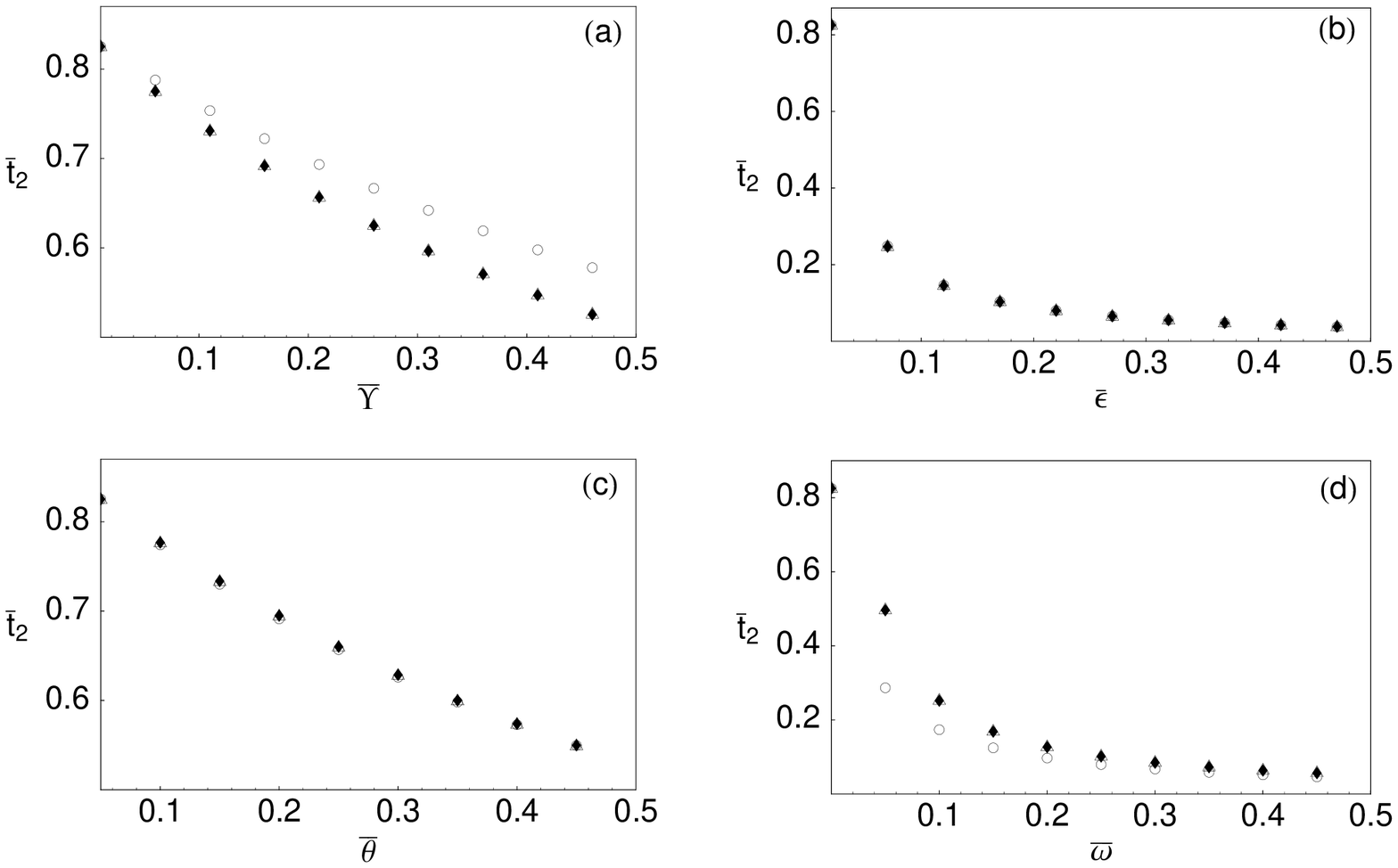}} 
\caption{Same as Fig.~\ref{t1} but for the characteristic time $\overline{t}_2$ on which the particle densities reach their diffusive quasi-equilibrium. We show a comparison of the characteristic time obtained from the order-of-magnitude estimate Eq.~(\ref{torder2lin}) (open circles) with that obtained from the numerical solution of the eigenvalue problem of Eq.~(\ref{eigen}) (black diamonds) and from the analytical approximation of Eq. ~(\ref{val2val3}) (open triangles).} 
\label{t2} 
\end{figure}

\begin{figure} 
\centering 
\resizebox{9cm}{9cm}{\includegraphics{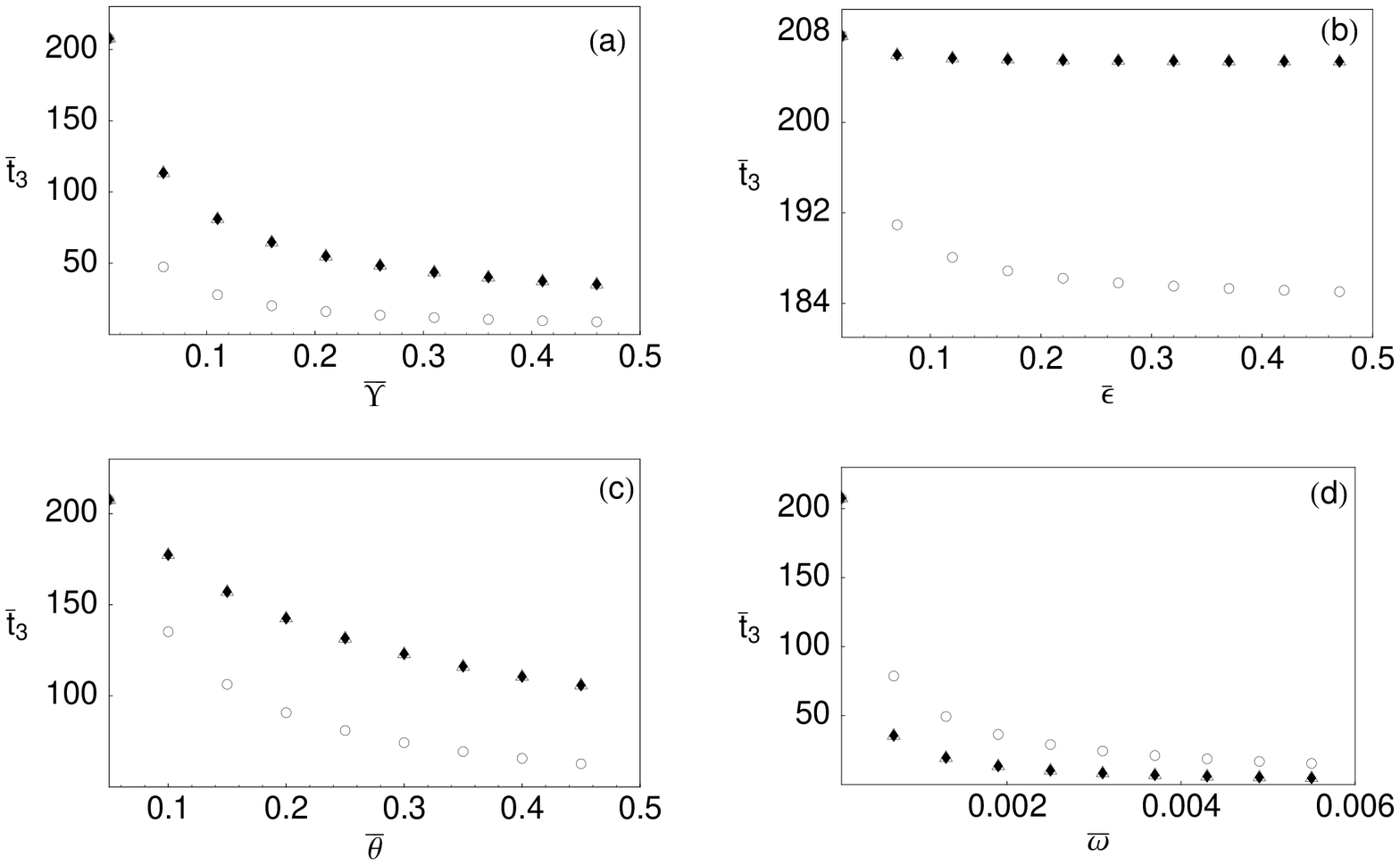}} 
\caption{Same as Fig.~\ref{t1} but for the characteristic time $\overline{t}_3$ on which the magnetic field evolves. We show a comparison of the characteristic time obtained from the order-of-magnitude estimate Eq.~(\ref{torder3lin}) (open circles) with that obtained from the numerical solution of the eigenvalue problem of Eq.~(\ref{eigen}) (black diamonds) and from the analytical approximation of Eq. ~(\ref{val2val3}) (open triangles).}
\label{t3} 
\end{figure}

We find normal-mode solutions for the dynamical variables in this linear system as 
 
\begin{equation} 
\label{nm1} 
\delta\overline{n}_B=\overline{a}_{\delta \overline{n}_B}(\overline{\eta}_m)\cos(l\pi\overline{x})\exp(-\overline{\eta}_m \overline t), 
\end{equation} 
 
\begin{equation} 
\label{nm2} 
\delta\overline{n}_c=\overline{a}_{\delta \overline{n}_c}(\overline{\eta}_m)\cos(l\pi\overline{x})\exp(-\overline{\eta}_m \overline t), 
\end{equation} 
 
\begin{equation} 
\label{nm3}  
\delta\overline{B}_z=\overline{a}_{\delta \overline{B}_z}(\overline{\eta}_m)\cos(l\pi\overline{x})\exp(-\overline{\eta}_m \overline t), 
\end{equation} 
where $m=1,2,3$ is an integer labeling each mode, $\overline \tau_m=1/\overline{\eta}_m$ is the decay time of each mode, $l$ is an integer measuring the mode wave-number, and $\overline{a}_{\delta \overline{p}}$ is the amplitude of the corresponding variable $\delta \overline{p}$ in each mode. We note that the function $\cos(l\pi\overline{x})$ satisfies the boundary conditions imposed in the previous section and that the characteristic length for spatial variations in this normal-mode solution is $L=d/(l\pi)$. We note that the linear decay times $\overline{\tau}_m$ correspond to the estimated characteristic timescales $\overline{t}_{m}$ in the general nonlinear regime [see Eqs. ~(\ref{torder1}), (\ref{torder2}), and (\ref{torder3})], and we use the same notation in the remainder of this paper. The normal-mode solution translates into an eigenvalue problem that can be solved either numerically or using analytical approximations. In Figs.~\ref{t1}-\ref{t3}, we compare the decay times of the normal modes obtained from an analytical approximation to the eigenvalue problem (open triangles; see Appendix \ref{ApA}) with those obtained from a numerical solution of the eigenvalue problem (black diamonds). We observe that in most cases there is good agreement between these two methods. Figures.~\ref{t1}(b) and \ref{t1}(d) show a discrepancy between these two methods, which is more noticeable for control parameters ($\overline \Upsilon$, $\overline \epsilon$, $\overline \theta$ and $\overline \omega$) of order of 1. This is expected since the analytical approximation to the eigenvalue problem should be accurate for control parameters significantly less than 1 (see Appendix \ref{ApA}). In Figs.~\ref{t1}-\ref{t3} we also plot the evolutionary timescales obtained from the order-of-magnitude estimates in Sect.~\ref{carsca} (open circles) [see Eqs. ~(\ref{torder1}), (\ref{torder2}) and (\ref{torder3})] with numerical coefficients inserted by hand to ensure reasonable agreement with the other, more precise determinations, 

\begin{equation} 
\label{torder1lin} 
\overline{t}_{1}= 1.1/(l\pi)^2, 
\end{equation}

\begin{eqnarray}
\label{torder2lin}
\overline {t}_{2}&=&0.71\left[(l\pi)^2 \overline \epsilon+\overline \theta \right.\nonumber\\
&&\left.+\left\{\frac{1}{7.5\overline \Upsilon}\left(\frac{1}{(l\pi)^2 \overline \epsilon}+\frac{1}{2.5\overline \theta}\right)\right\}^{-1}+0.81(l\pi)^2 \overline \omega \right]^{-1},
\end{eqnarray}

\begin{eqnarray}
\label{torder3lin} 
\overline {t}_{3}&=&2.3\left[\left[(l\pi)^2 \overline \epsilon+\overline \theta\right]^{-1}\right.\nonumber\\
&&\left.+\left[\left\{\frac{1}{7.5\overline \Upsilon}\left(\frac{1}{(l\pi)^2 \overline \epsilon}+\frac{1}{2.5\overline \theta}\right)\right\}^{-1}+0.81(l\pi)^2 \overline \omega \right]^{-1}\right].
\end{eqnarray}

The fact that this agreement can be reached using coefficients that do not differ significantly from unity (see Eqs. ~(\ref{torder1lin}), (\ref{torder2lin}) and (\ref{torder3lin})) corroborates the order-of-magnitude estimates in the framework of the linear-mode regime. 

\subsection{Finite difference evolution of a linear magnetic-field perturbation}
\label{finline} 
We developed a numerical code based on a finite-difference algorithm that solves the nonlinear set of Eqs.~(\ref{magbar}), (\ref{nbnorm}), and (\ref{ncnorm}) (see Appendix \ref{ApB}). In this section, we use that code to simulate the evolution of a linear magnetic-field perturbation of the form given by Eq.~(\ref{bpert1}), as well as the associated particle-density evolution. In this task, we aim to achieve two main goals: 
\begin{enumerate} 
\item To verify the quasi-equilibrium states that occur during the evolution. 
\item To compare the linear normal-mode solution given by Eqs.~(\ref{nm1}), (\ref{nm2}), and(\ref{nm3}) with the solution obtained by solving numerically the set of Eqs.~(\ref{magbar}), (\ref{nbnorm})\large and (\ref{ncnorm})(see Appendix \ref{ApB}), to measure the precision of the numerical code. 
\end{enumerate} 
  
\begin{figure} 
\resizebox{7cm}{5cm}{\includegraphics{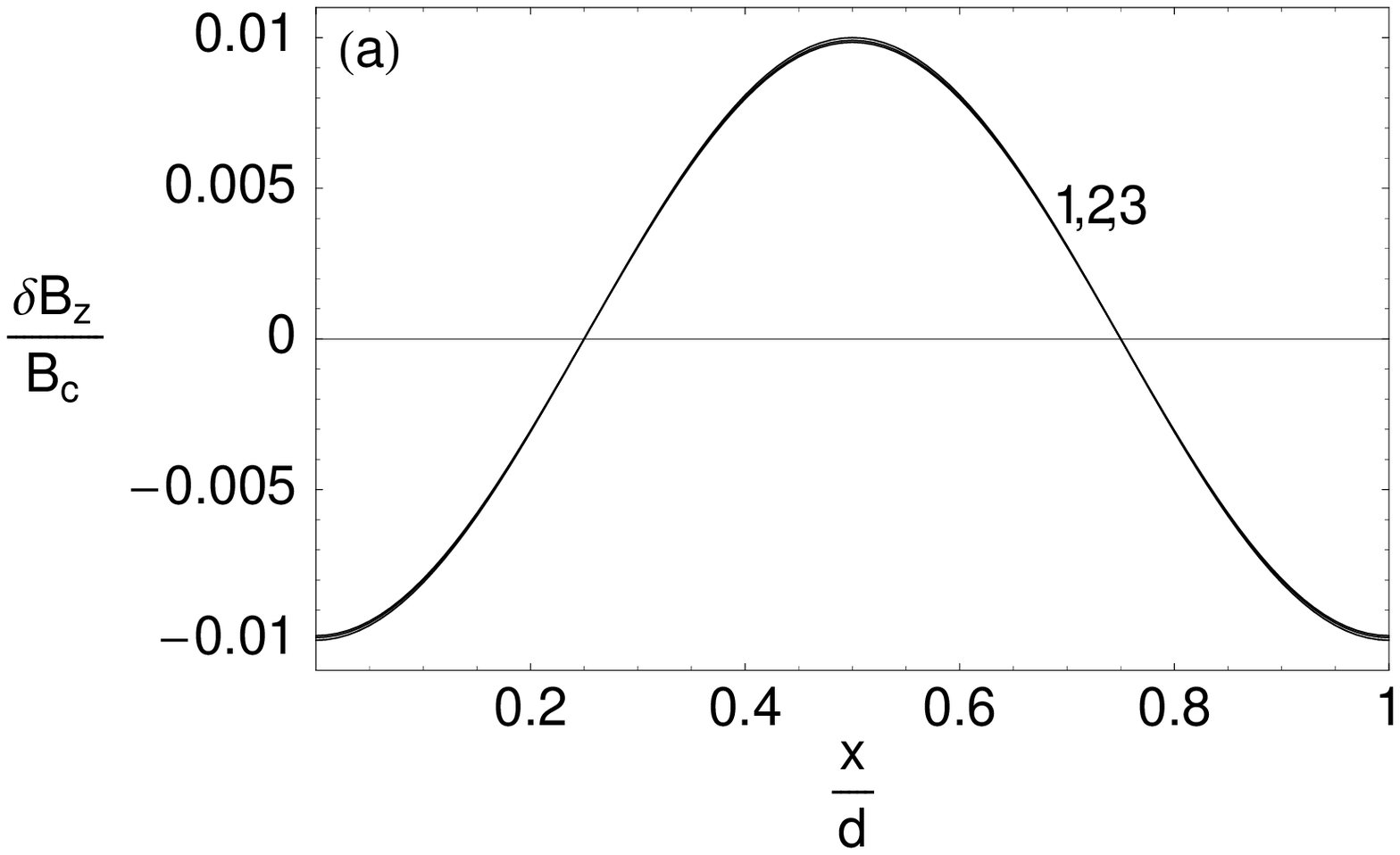}} 
\resizebox{7cm}{5cm}{\includegraphics{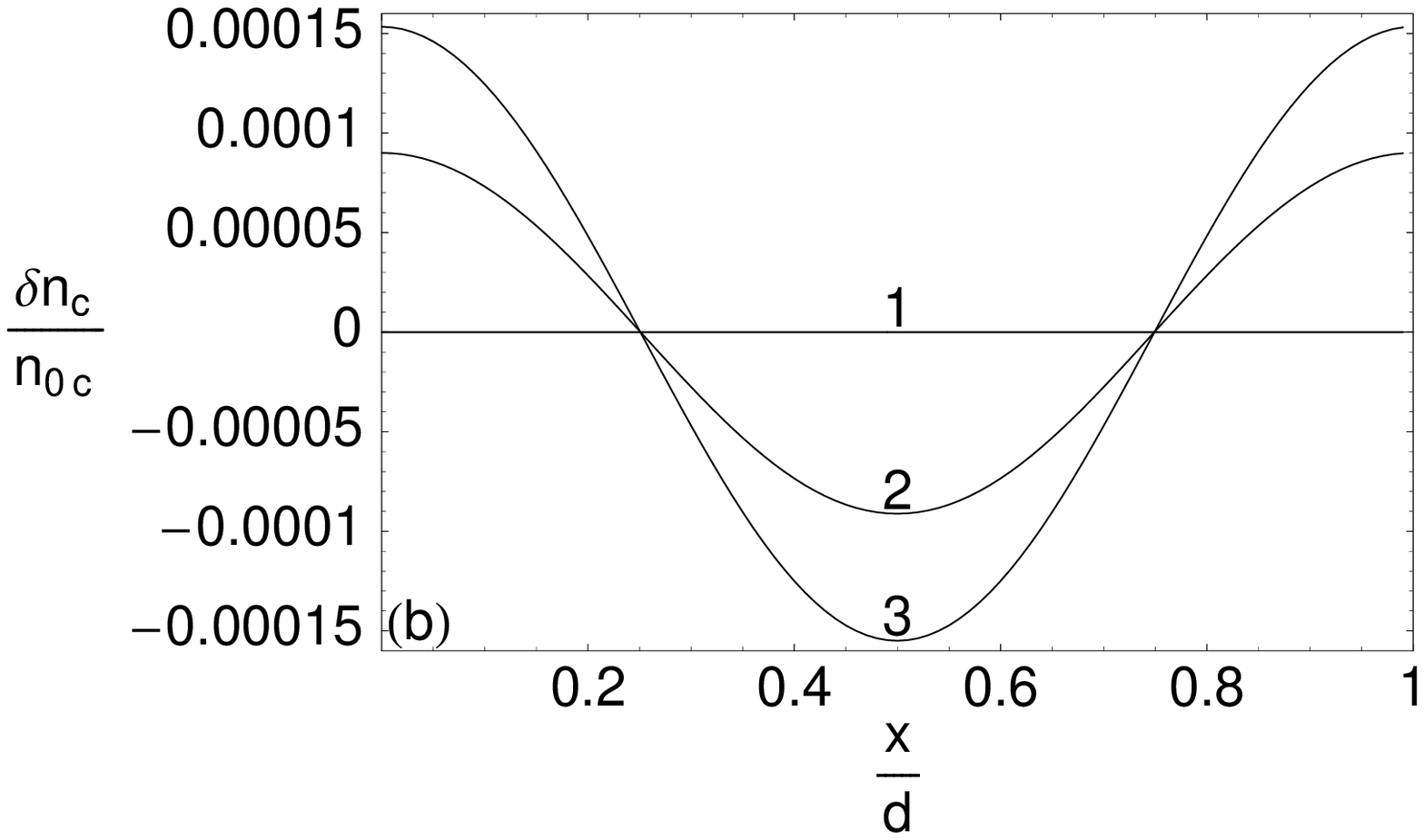}} 
\centering 
\resizebox{7cm}{5cm}{\includegraphics{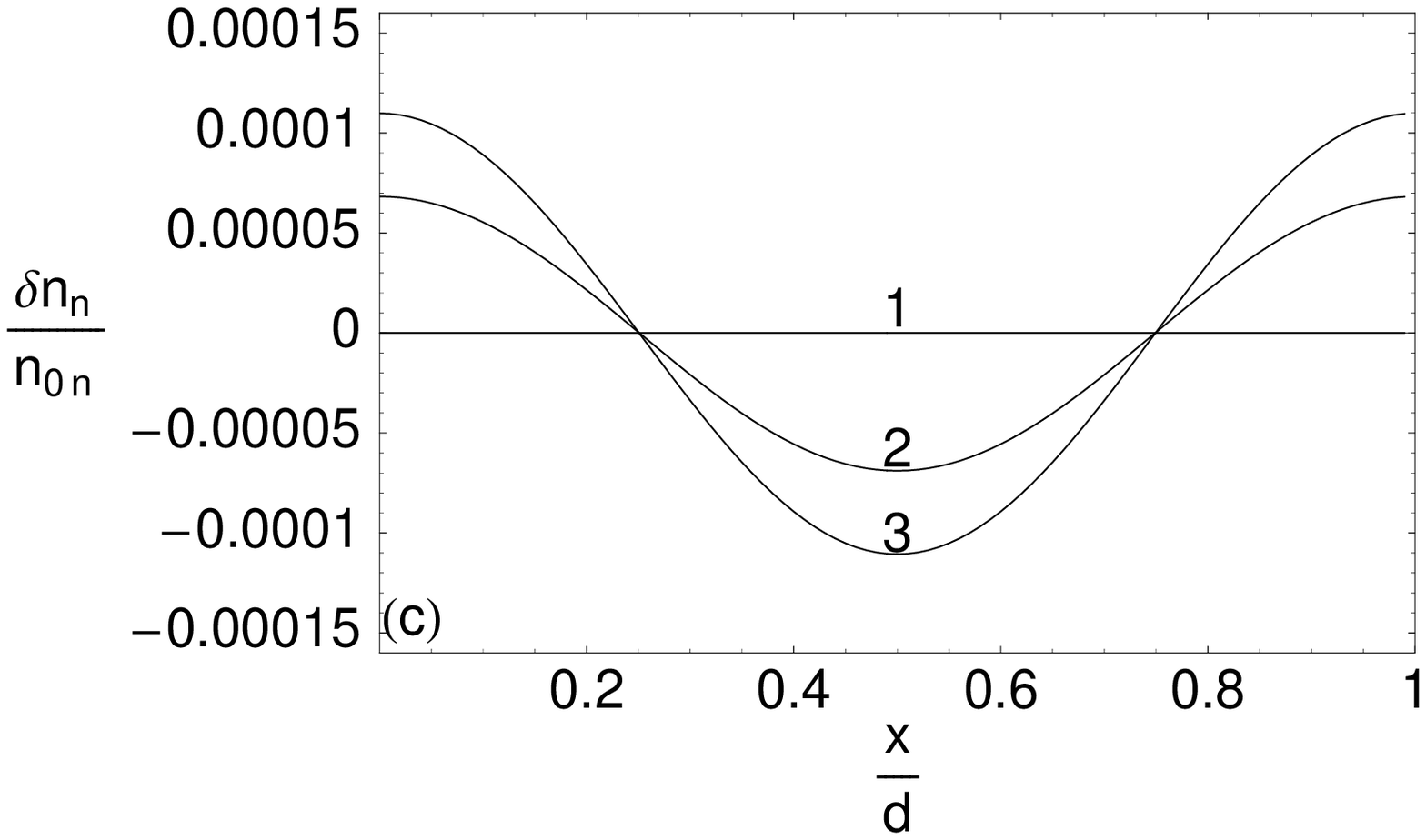}} 
\caption{Evolution of a magnetic field of the form $B_z(x,t)=B_c+\delta B_z(x,t),$ where $B_c$ is a constant and $\delta B_z$ is a small perturbation  (i.e., $|\delta B_z|\ll B_c$). The associated charged-particle and neutrons density perturbations around a homogeneous background star are $\delta n_c(x,t)$ and $\delta n_n(x,t)$ respectively, with the neutron and charged-particle number densities in the background defined by $n_{0n},$ $n_{0c},$ and $d$ the total length of the system. This figure shows the evolution during the first timescale $t_{1}$ on which the system reaches the magnetohydrostatic quasi-equilibrium. The time variable is normalized as in Figs.~\ref{t1},~\ref{t2} and \ref{t3}. We have used in this simulation $n_{0n}=9.8 \times 10^{37} \mathrm{cm}^{-3}$ and $n_{0c}/n_{0n}=4.0\times 10^{-2}.$ We set the control parameters of the different physical processes to: $\overline \omega=0,$ $\overline \Upsilon=0.2,$ $\overline \theta=0.1,$ and $\overline \epsilon =0.01,$ which gives $\overline{t}_{1}=0.028.$ (a) Magnetic field perturbation at different instants labeled with progressive numbers starting with (1) for the initial condition at $\overline t=0.$ The initial magnetic perturbation is $\delta B_z(x, 0)=B_c\overline A \cos(l\pi x/d),$ with $\overline A=-0.010$ and $l=2.$ while the initial particle density perturbations are $\delta {n}_c(x, 0)=\delta {n}_n(x, 0)=0.$ The other instants are(2): $\overline t=0.014,$  (3): $\overline t=0.028.$ (b) Charged-particle density perturbations at the same instants as in panel (a). (c) Neutron density perturbations at the same instants as in panel (a).}
\label{tmhd} 
\end{figure}

\begin{figure} 
\resizebox{7cm}{5cm}{\includegraphics{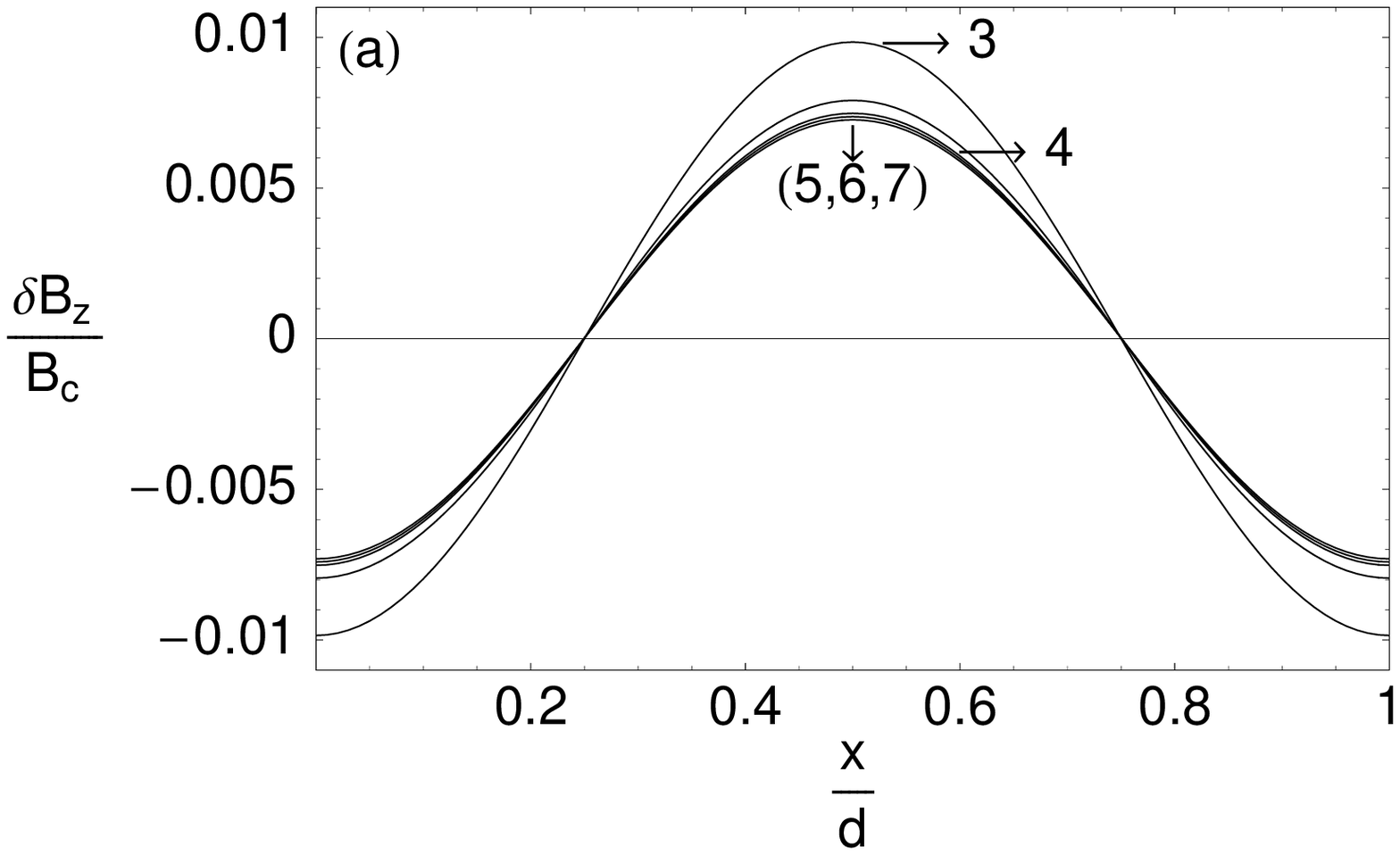}} 
\resizebox{7cm}{5cm}{\includegraphics{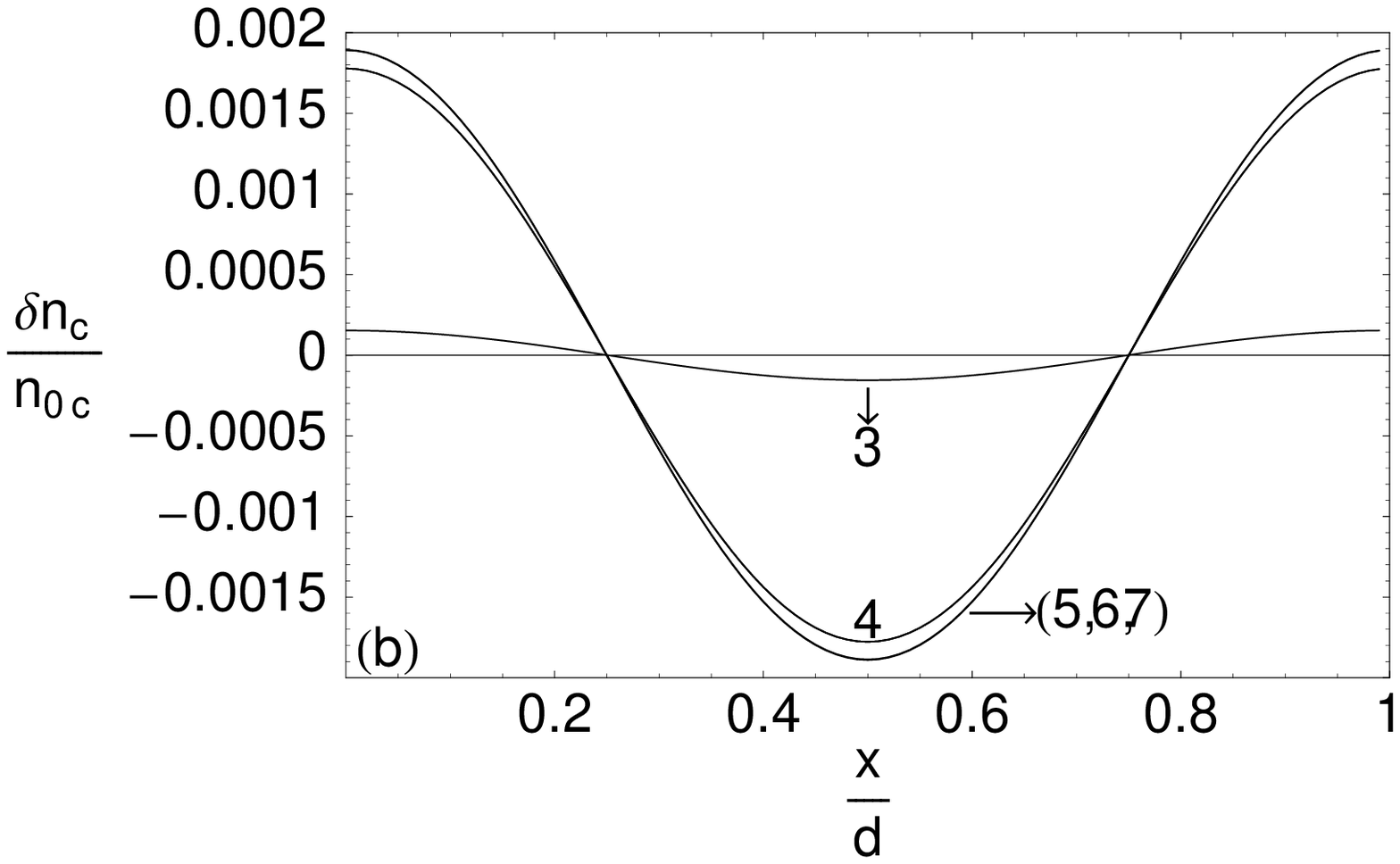}} 
\centering 
\resizebox{7cm}{5cm}{\includegraphics{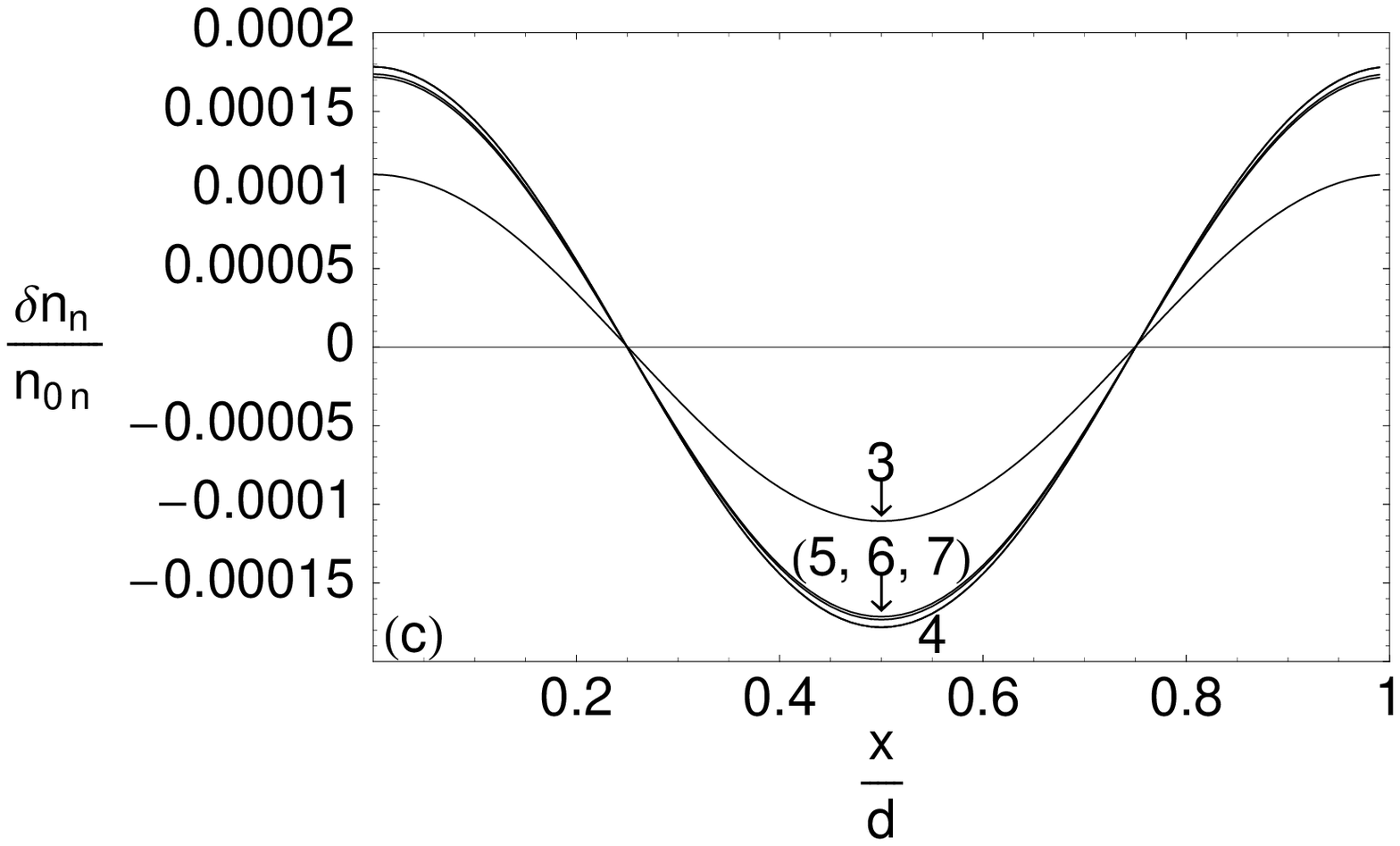}} 
\caption{Same parameters and normalization conventions than Fig.~\ref{tmhd} but showing the evolution during the second timescale $\overline t_{2}$ on which the particles densities evolve. For these parameters $\overline{t}_{2}=1.2.$ (a) Magnetic field perturbation at different instants. The instant labels are: ($3$): $\overline t=0.028,$ ($4$): $\overline t=2.4,$ ($5$): $\overline t=3.8,$ ($6$): $\overline t=4.2,$ ($7$): $\overline t=4.7.$ (b) Charged-particle density perturbations at the same instants as in panel (a). (c) Neutron density perturbations at the same instants as in panel (a).}
\label{tdif} 
\end{figure}

\begin{figure} 
\resizebox{7cm}{5cm}{\includegraphics{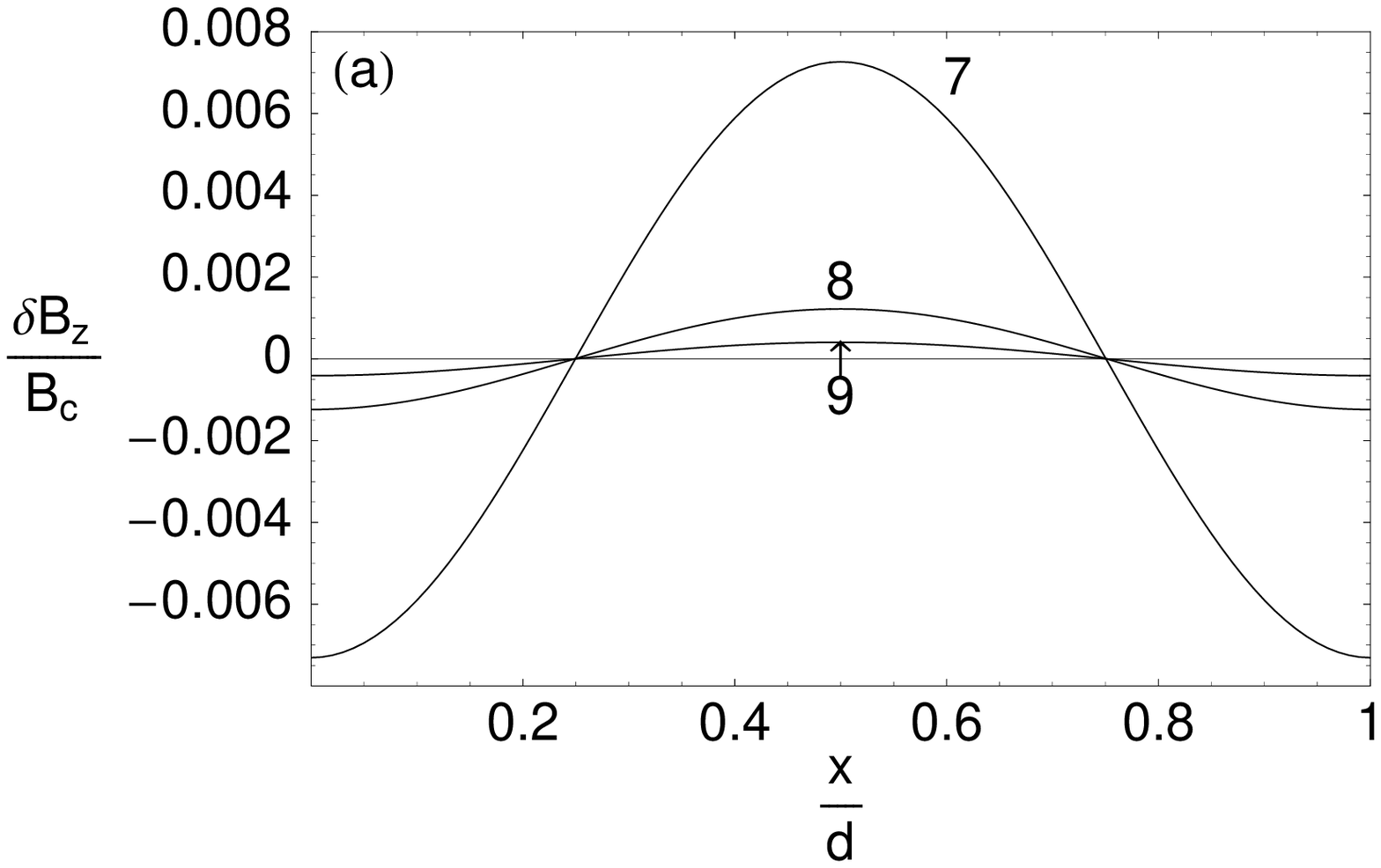}} 
\resizebox{7cm}{5cm}{\includegraphics{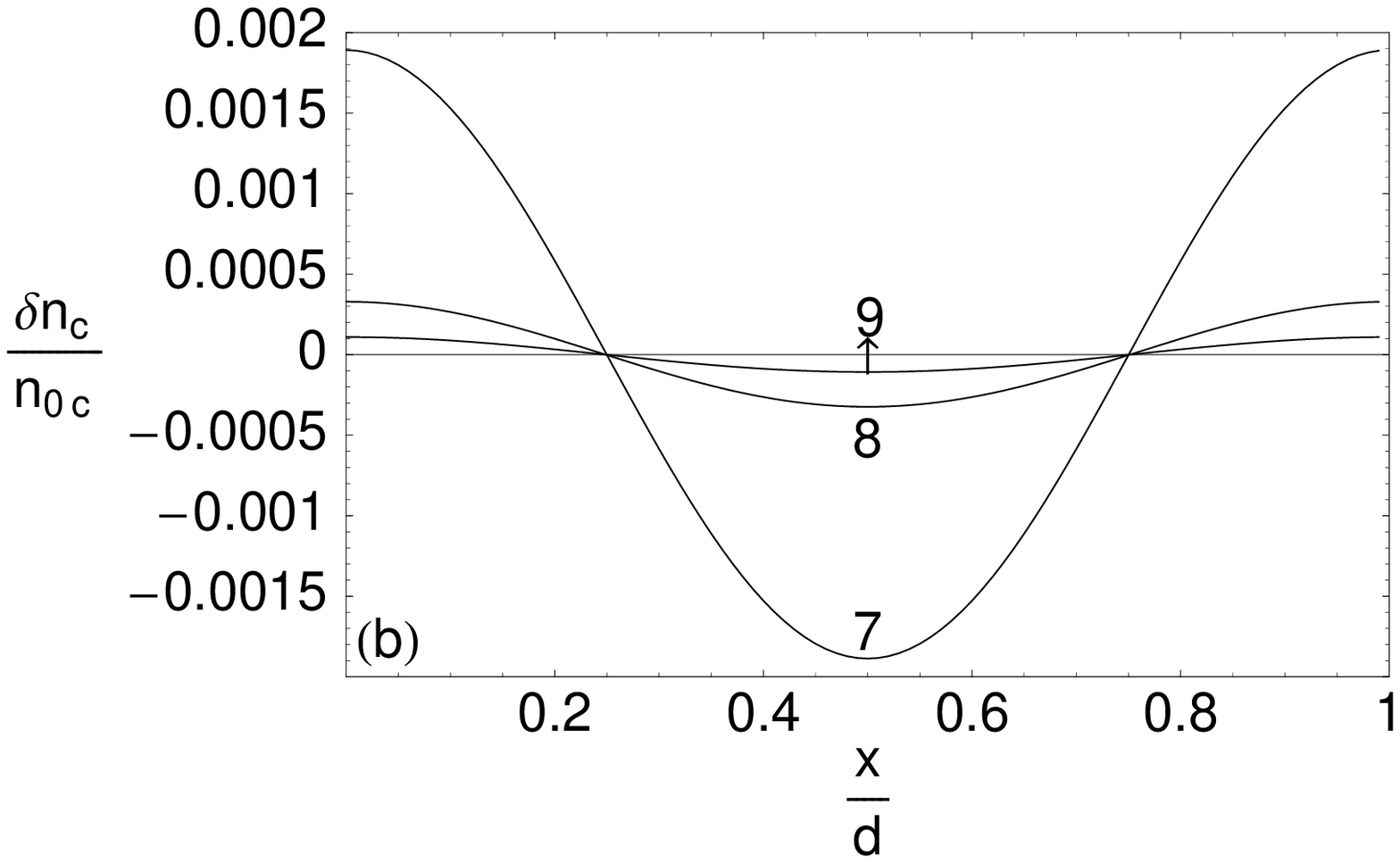}} 
\centering 
\resizebox{7cm}{5cm}{\includegraphics{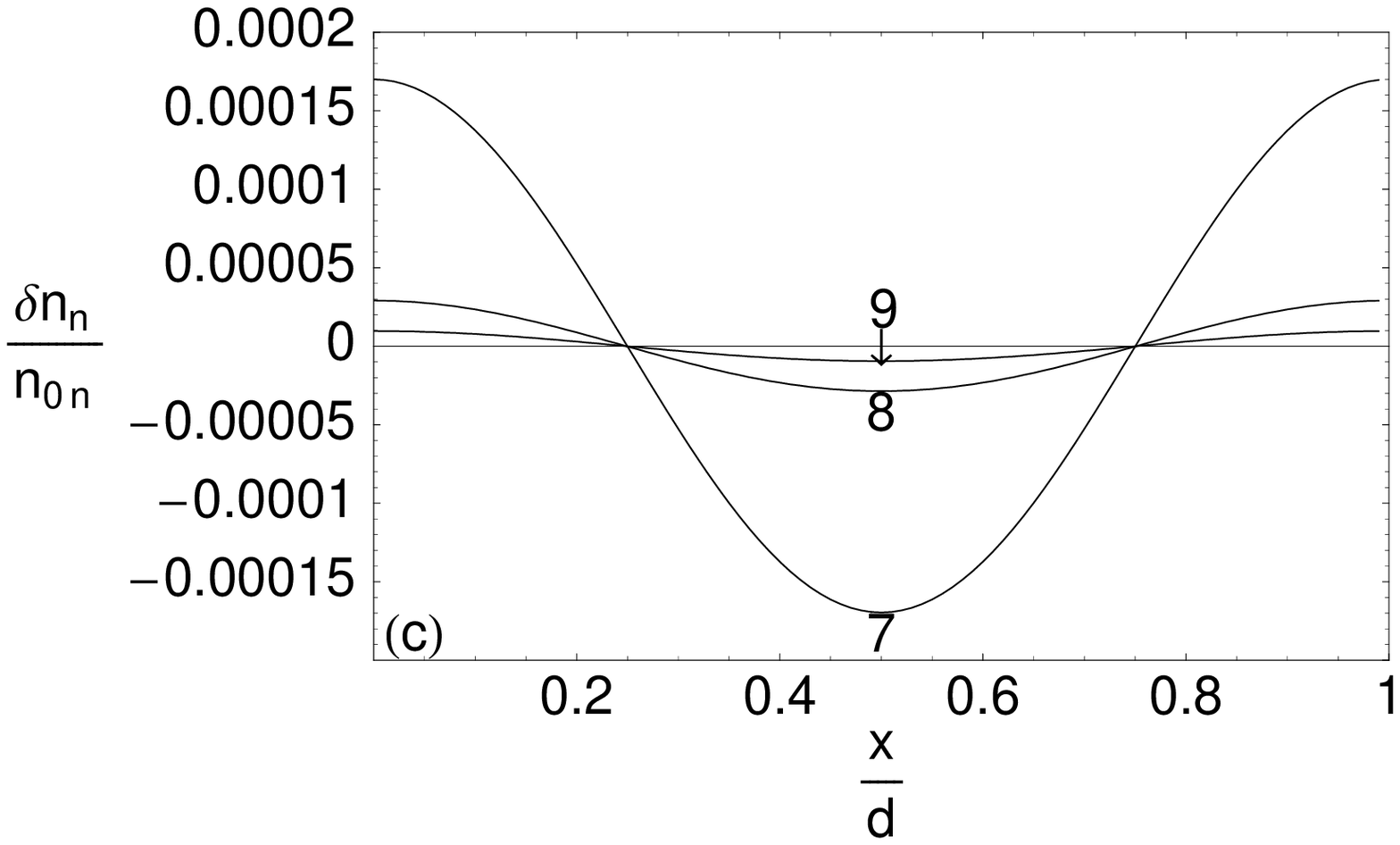}} 
\caption{Same parameters and normalization conventions than Fig.~\ref{tmhd} but showing the evolution during the third timescale $\overline t_{3}$ on which the magnetic evolves significantly. For these parameters $\overline{t}_{3}=39.0.$ (a) Magnetic field perturbation at different instants. The instant labels are: ($7$): $\overline t=4.7,$ ($8$): $\overline t=27.0,$ ($9$): $\overline t=116.0,$ (b) Charged-particle density perturbations at the same instants as in panel (a). (c) Neutron density perturbations at the same instants as in panel (a).} 
\label{tchem} 
\end{figure}

In our numerical simulation, we set the initial condition to correspond to a magnetic field of the form given by Eq.~(\ref{bpert1}), with $\delta \overline B_z(\overline x, 0)=\overline A \cos(l\pi\overline{x})$, $\overline A=-0.010$ and $l=2$. We set the initial particle density perturbations to be $\delta \overline{n}_c(\overline x, 0)=\delta \overline{n}_n(\overline x, 0)=0$. In what follows, we neglect the ohmic diffusion, therefore, we expect that $\overline{t}_{1}=\overline{t}_{MHS}$, $\overline{t}_{2}=\overline{t}_{\delta n_c}$, and  $\overline{t}_{3}=\overline{t}_{B}$. We use the parameters: $\overline \omega=0$, $\overline \Upsilon=0.20$, $\overline \theta=0.10$, $\overline \epsilon =0.010$. We note we are in a regime in which ambipolar diffusion occurs more rapidly than the weak interaction processes ($\overline{t}_{drag}\approx 2.5$ and $\overline{t}_{weak}\approx 10$). From a numerical solution of the eigen-value problem [see Eq.~(\ref{eigen})], we obtain $\overline{t}_{1}=0.028$, $\overline{t}_{2}=1.2$, and  $\overline{t}_{3}=39.0$. In Fig.~\ref{tmhd}, we show the system's evolution during the first timescale $\overline{t}_1$, which is the timescale over which the system can reaches magnetohydrostatic quasi-equilibrium. We label the different instants of this evolution with progressive numbers as described in the figure. This magnetic perturbation perturbs the hydrostatic quasi-equilibrium of the homogeneous background star; thus particles are therefore compelled to move in order to compensate for this imbalance, and reach magnetohydrostatic quasi-equilibrium in a short timescale of the order of $\overline{t}_1$. In Fig.~\ref{tmhd} (a) we observe the evolution of the magnetic-field perturbation during this short timescale, and that it has not evolved significantly. This is expected since the magnetic pressure is small compared with the background-degeneracy pressure, and can therefore induce only a small relative density perturbation in the particles (see Figs.~\ref{tmhd} (b) and \ref{tmhd} (c)). We can also see in Figs.~\ref{tmhd} (b) and \ref{tmhd} (c) that during this short timescale $\delta{n}_c/{n}_{0c} \approx \delta{n}_n/{n}_{0n}$, as expected from Eq.~(\ref{mhdc}). In Fig.~\ref{tdif}, we show the system's evolution during the second timescale  $\overline{t}_2$, on which the diffusive quasi-equilibrium of the charged particles is attained. At the end of this timescale, $\delta{n}_n/{n}_{0n}\sim 0$, and $\delta{n}_c/{n}_{0c}$ has grown significantly to balance the magnetic pressure. Towards the end of this timescale, the separation between the lines is narrower, showing that the charged particles have reached their diffusive quasi-equilibrium. In Fig.~\ref{tchem}, we show the system's evolution during the third timescale  $\overline{t}_3$, on which the chemical quasi-equilibrium is restored and the magnetic perturbation decays. At the end of this timescale, the density perturbations as well as the magnetic-field perturbation have decreased substantially.  

On the other hand, we note that a linear magnetic-field perturbation can be written in general, as a superposition of the linear normal modes given by Eq.~(\ref{nm3}) as  
 
\begin{equation} 
\label{sup} 
\delta \overline{B}_z(\overline x, \overline t)=\sum_{m=1}^3 \overline{D}_m \overline a_{\delta \overline{B}_z}(\overline \eta_m) \cos(l\pi\overline{x}) \exp(-\overline \eta_m \overline t). 
\end{equation} 
where $\overline \eta_m = 1/ \overline t_m$ and the coefficients $\overline{D}_m$ are calculated from the initial conditions $\delta \overline{B}_z(\overline x,0)$, $\delta \overline{n}_c(\overline x, 0)$, and $\delta \overline{n}_n(\overline x, 0)$. For the initial conditions of this simulation and from the numerical solution of Eq.~(\ref{eigen}), we derive that $\overline a_{\delta\overline{B}_z}(\overline \eta_1)=-0.0076$, $\overline a_{\delta\overline{B}_z}(\overline \eta_2)=0.10$, $\overline a_{\delta \overline{B}_z}(\overline \eta_3)=0.22$, $\overline{D}_1=0.023$, $\overline{D}_2=-0.016$ and $\overline{D}_3=-0.037$. We Note that, for $\overline{t} \gg \overline{t}_2$,  
 
\begin{equation} 
\label{asymmm} 
\delta \overline{B}_z(\overline x, \overline t) \approx \overline{D}_3 \overline a_{\delta\overline{B}_z}(\overline \eta_3) \cos(l\pi\overline{x}) \exp(-\overline \eta_3 \overline t). 
\end{equation} 
 
\begin{figure} 
\centering 
\resizebox{9cm}{7cm}{\includegraphics{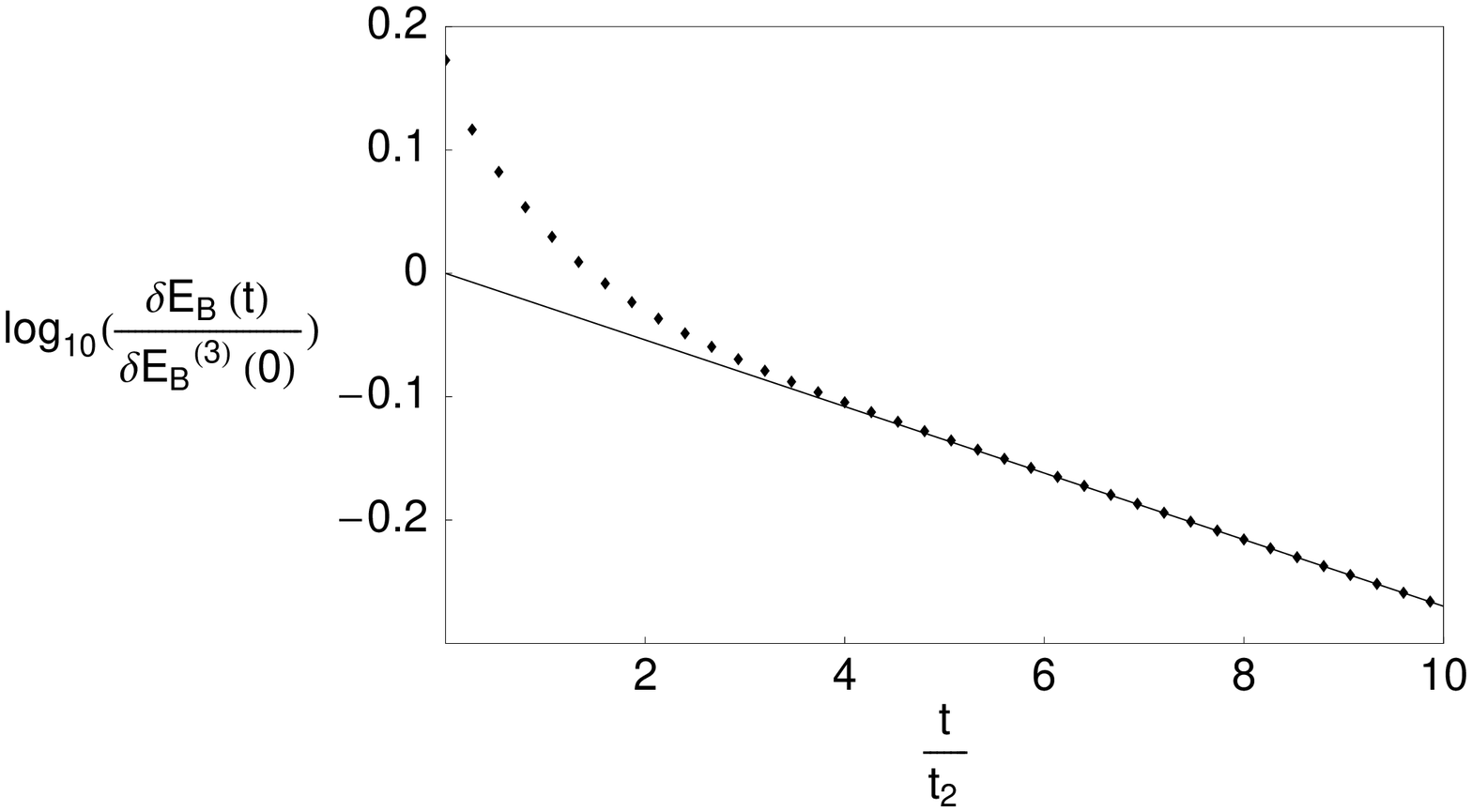}} 
\caption{Decay of the total magnetic energy contained in the perturbation of Fig.~\ref{tmhd} (a), Fig.~\ref{tdif} (a), and Fig.~\ref{tchem} (a) $\delta {E}_B(t)= (1/8\pi) \int_0^d (\delta{B}_z)^2 dx$ (points). We compare this solution with the normal mode solution in the asymptotic limit of Eq.~(\ref{asymagneenergl}) (full line). The quantity $\delta {E}_B^{(3)}(0)$ is the initial energy contained in the third mode whose decay time is $t_3.$ The time variable is normalized respect to the decay time of the second mode $t_2.$ For the parameters of this simulation $t_3/t_2=32.5.$} 
\label{linen} 
\end{figure}

On the other hand, the total magnetic energy associated with this perturbation in physical units is 
 
\begin{equation} 
\label{magneenergl} 
\delta {E}_B(t)= \int_0^d \frac{(\delta{B}_z)^2}{8\pi} dx. 
\end{equation} 
Thus, in the asymptotic limit of Eq.~(\ref{asymmm}), 
 
\begin{equation} 
\label{asymagneenergl} 
\log(\delta {E}_B(t))\approx  \log(\delta {E}_B^{(3)}(0))-2 \log(e) \frac{t}{t_3}. 
\end{equation} 
where $\delta {E}_B^{(3)}(0)=\left(D_3 a_{\delta B_z}(\eta_3)\right)^2 d/(16\pi)$ is the initial energy contained in the third mode. To measure the precision of our numerical code, in Fig.~\ref{linen} we compare the evolution of the magnetic-energy perturbation obtained from the simulation with the asymptotic limit of Eq.~(\ref{asymagneenergl}). In Fig.~\ref{linen}, we see that there is good agreement between our numerical result and the asymptotic limit of Eq.~(\ref{asymagneenergl}).

\subsection{Finite difference evolution of a non-linear magnetic field} 
\label{finnoli}
 
Using our numerical code, we simulate the evolution of a non-linear magnetic field, by applying the set of eqs. (\ref{magbar}), (\ref{nbnorm}), and (\ref{ncnorm}) without any linearization of $\overline{B}_z$. The main aims of this section are:  
 
\begin{enumerate} 
\item To compare the estimate obtained using Eq.~(\ref{tbch}) with the result of the numerical code in the non-linear regime. 
\item To verify the conservation laws given by  Eq.~(\ref{conserv}) for the magnetic flux and baryon number, which provide a measure of the right performance of our numerical code. 
\end{enumerate} 
 
Defining the initial conditions of our simulation, we set the particle densities, $\delta \overline{n}_c(\overline x, 0)=\delta \overline{n}_n(\overline x, 0)=0$, for an initial Gaussian magnetic field profile given by 
 
\begin{equation} 
\label{nonlineraB} 
\overline{B}_{z}(\overline x,0)=\exp(-s^2(\overline x-\overline x_0 )^2). 
\end{equation} 
 
We normalize the magnetic field in Eq.~(\ref{nonlineraB}) to its maximum value $B_z^{max}$. The parameter $s$ in Eq.~(\ref{nonlineraB}) is related to its standard deviation $p$ by $s^2=d^2/2p^2$, and $\overline x_0 = x_0/d$ is the center of the Gaussian function. The characteristic spatial variation in the Gaussian function can be estimated by the width $w$ of the Gaussian function at its half-height, which is given by $w\approx 2.355 p$; we can therefore approximate the characteristic length over which the magnetic force varies to be $w \approx L \approx 2.355d/\sqrt{2}s$. In our analysis, we again neglect ohmic diffusion. To estimate the characteristic decay time of the magnetic field in Eq.~(\ref{nonlineraB}), we use as a first approximation the characteristic length of the magnetic force in Eq.~(\ref{tbch}) as the width of the initial gaussian, therefore, 
 
\begin{equation} 
\label{tbchnol} 
\overline{t}_B\sim \frac{1}{\overline \Upsilon}\left(\frac{2.355^2}{2s^2}\frac{1}{\overline \epsilon}+\frac{1}{\overline \theta}\right). 
\end{equation} 

\begin{figure} 
\resizebox{7cm}{5cm}{\includegraphics{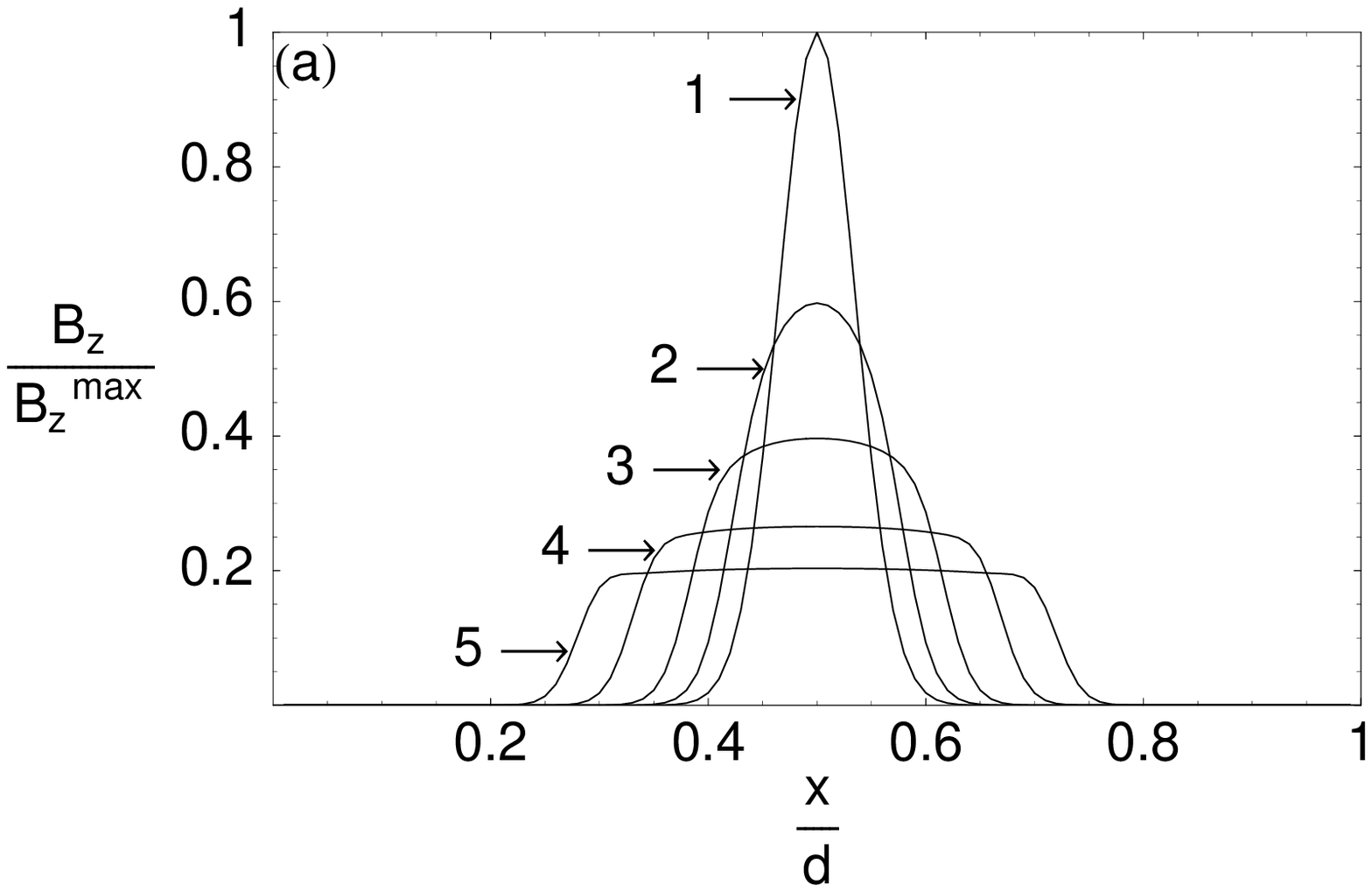}} 
\resizebox{7cm}{5cm}{\includegraphics{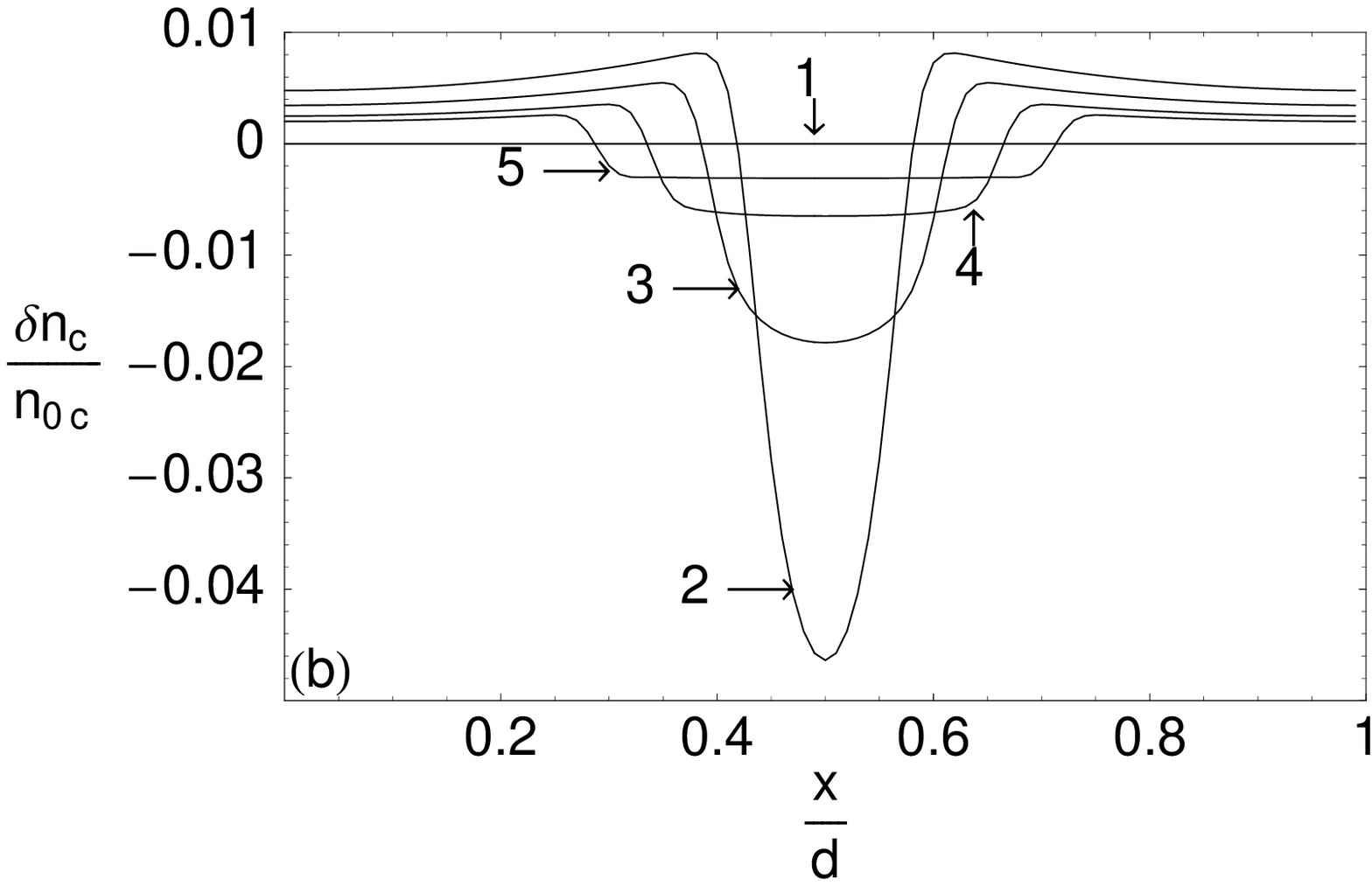}} 
\centering 
\resizebox{7cm}{5cm}{\includegraphics{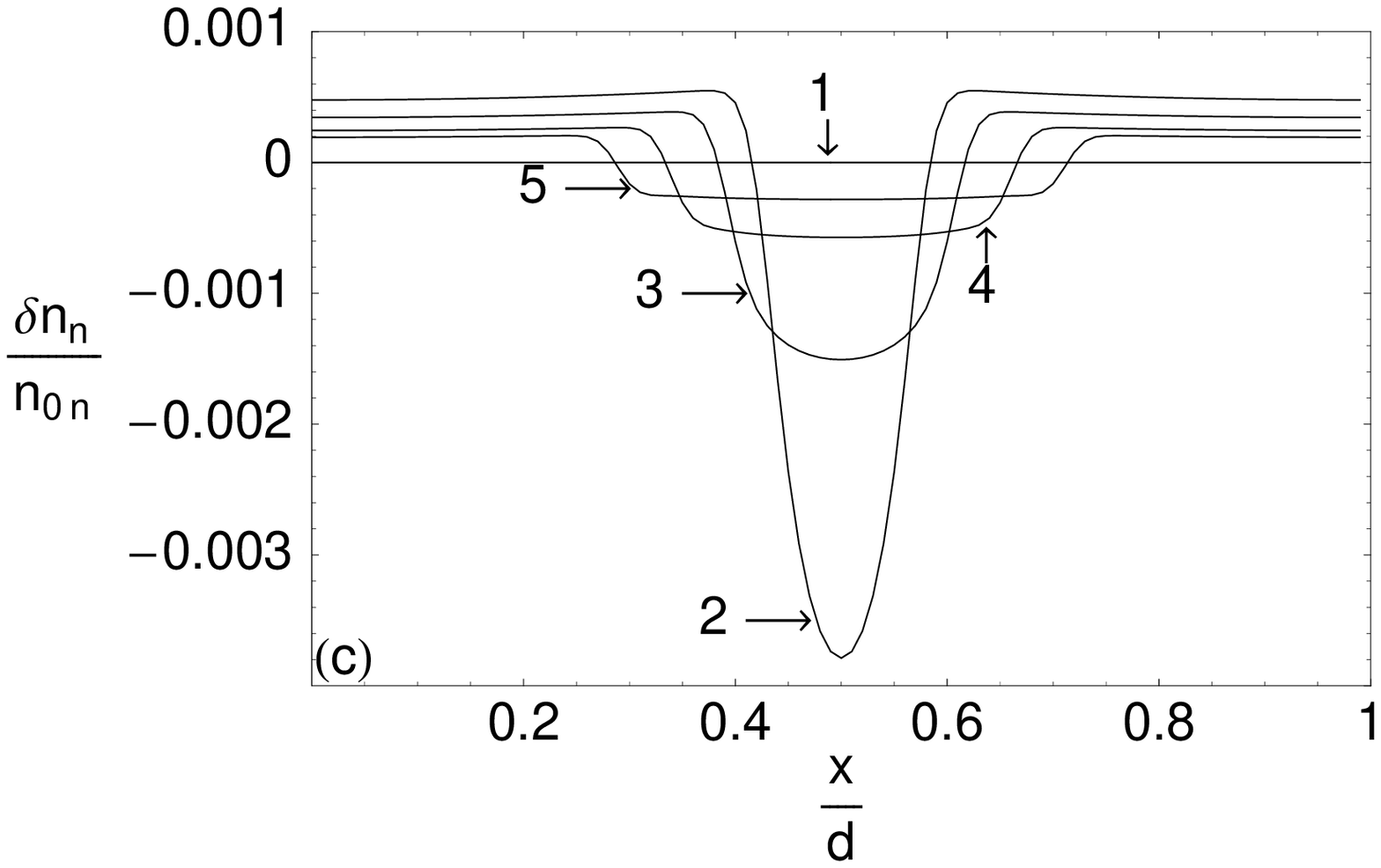}} 
\caption{Evolution of an initial Gaussian magnetic field profile given by ${B}_{z}(x,0)=B_z^{max}\exp(-(20/d)^2(x-(d/2))^2)$ as well the associated particle density perturbations. We have used the same background parameters and time normalization convention as in Fig.~\ref{tmhd}. The initial condition in the particle densities is such that they are uniformly distributed in the background, thus $\delta{n}_n({x},0)=\delta{n}_c({x},0)=0.$ We set the parameters controling the different physical processes as: $\overline \omega=0$, $\overline \epsilon=0.01$, $\overline \theta=0.1$, and $\overline \Upsilon=0.2$.  The characteristic decay time over which the magnetic field decays significantly is $\overline{t}_B\approx 53.5$. (a) Evolution of the magnetic field for different instants labeled with progresive numbers, $(1):\overline{t}=0$.  $(2):\overline{t}=54$, $(3):\overline{t}=180$, $(4):\overline{t}=500$, $(5):\overline{t}=1000$. (b) Evolution of the charged-particle density perturbations for the same instants as in panel (a). (c) Evolution of the neutron density perturbations for the same instants as in panel (a).} 
\label{gauss} 
\end{figure}

In our numerical analysis, we assume that $s=20.0$, $\overline x_0=0.5$, $\overline \omega=0$, $\overline \epsilon=0.01$, $\overline \theta=0.1$, and $\overline \Upsilon=0.2$. If we substitute these parameters in Eq.~(\ref{tbchnol}), we obtain a characteristic decay time $\overline{t}_B \approx 53.5$. In Fig.~\ref{gauss} (a), we see the evolution of the initial magnetic field given by Eq.~(\ref{nonlineraB}), and that at times close to $\overline{t}_B$ the maximum of the magnetic field has decayed to $60\%$ of its initial amplitude. On the other hand, Fig.~\ref{gauss} (b) shows the evolution of the charged-particle density perturbations, and Fig.~\ref{gauss} (c) shows the evolution of the neutron density perturbations.

We verified the conservation law for the dimensionless baryon number $\overline{N}_{B}(\overline t)=\overline{N}_{0B}+\delta \overline{N}_B(\overline t)$, where $\overline{N}_{0B}=1.04$ was the backgound baryon number. According to the initial conditions $\delta \overline{N}_B(0)=0,$ which implies that the conservation law requires $\overline{N}_{B}(\overline t)=\overline{N}_B(0)=\overline{N}_{0B}.$ The calculated values of the baryon number in the simulation time steps show a maximum percentage error of $2.0\times10^{-5}~\%$ with respect to $\overline{N}_{0B}$. Regarding the magnetic flux, we found exact conservation with respect to the initial magnetic flux $\overline{\Phi}_B(0)$ down to 7-digit precision. 
  
\section{Summary and Conclusions}
\label{conc} 
 
We have studied the long-term evolution of the magnetic field and the densities of neutrons and charged particles (protons and electrons in the interior of a neutron star, using a multi-fluid model with a simplified geometry in which the magnetic field points in one Cartesian direction and varies along an orthogonal direction. We found a set of three non-linear partial differential equations of first order in time describing this evolution, and we analyzed them in three different ways: (i) estimating evolutionary timescales directly from the equations, guided by physical intuition; (ii) a normal-mode analysis of the equations in the limit of a nearly uniform system; and (iii) a finite-difference numerical integration of the equations. In this section, we summarize the main results of each one of these approaches and present the main conclusions of this work.  
  
\subsection{Evolutionary timescales} 
\label{conc1}
 
The three partial differential equations of our model constitute a dynamical system of three independent variables (magnetic-field, charged-particle, and neutron density perturbations). We identified three characteristic evolutionary timescales on which the system approaches successive quasi-equilibrium states. In the following, we summarize the basic physical processes governing the evolution and the estimates of each timescale.
 
\subsubsection{Timescale to achieve magnetohydrostatic quasi-equilibrium} 
 \label{conc11}
During the early stages of a neutron star's life, the Lorentz force moves the bulk stellar fluid, inducing perturbations in its pressure. In this short time, all particles move together as a single fluid with the same bulk velocity. There is also not enough time for weak interactions between particles (beta decays) to operate, so the composition is frozen. The system evolves until it reaches a magnetohydrodystatic quasi-equilibrium state, in which the Lorentz and the fluid forces are close to balancing. Neutron stars with no rotation or convection reach this quasi-equilibrium state in a short time, not much longer than the Alfv\'en time, which for typical magnetar core parameters scales as 
 
\begin{equation} 
\label{tafr} 
t_{Alfven}=5.7~\times 10^{-2}~ R_{6}~B_{15}^{-1}~s, 
\end{equation}  
where $R_{6}\equiv R/(10^6~cm)$ denotes the radius of the star in units of $10^6~cm$ and $B_{15}\equiv B_z/(10^{15}~G)$ the magnetic field in units of $10^{15}~G.$ On timescales far longer than the Alfv\'en time, relative movements between the different species of particles in the star as well as weak interactions between them become relevant (GR-92). Therefore, a description of the system during these stages requires a multi-fluid theory. In this paper, we develop a model to study the decay of the magnetic field induced by these long-term mechanisms. The Alfv\'en time is far shorter than the timescales of these long-term processes (see Sects.~\ref{conc12} and \ref{conc13}), which are of the order of years or much more. A numerical code simulating the evolution during the Alfv\'en timescale would require a time step many orders of magnitude shorter than that required to simulate the long-term evolution in a computational time not prohibitively long. In our model, we overcome this difficulty by replacing the short-term dynamics by a ficticious friction force term acting on the neutrons (the most abundant species). Its strength is controlled by an artificial parameter $\alpha$, chosen so that the timescale to reach  magnetohydrostatic quasi-equilibrium is long enough for the numerical code to be able to deal with it (and therefore much longer than the Alfv\'en time), but shorter than the timescales of the long-term processes that we are interested in modeling. 
 
\subsubsection{Timescale for the evolution of the particle densities}
\label{conc12}  
 
On timescales far longer than required to reach magnetohydrostatic quasi-equilibrium, the neutrons and charged particles move relative to each other with different velocities and are affected by collisional drag. Also, weak interactions (beta decays) convert particles from one species into another, erasing departures from chemical quasi-equilibrium among neutrons, protons, and electrons caused by the induced particle density perturbations. Both processes contribute to the evolution of the particle densities. In estimating the timescale for this evolution, it is useful to distinguish the regimes in which each one of these processes is dominant. If the collisions between the charged particles and the neutrons are rare and the weak interactions are slow, the charged particles move easily relative to the neutrons (ambipolar diffusion) allowing the particles to reach a diffusive quasi-equilibrium state (in which gravitational, electromagnetic, and pressure forces are closely balanced) but staying avoid chemical quasi-equilibrium. The diffusive quasi-equilibrium is reached in a timescale controlled by the collision rate between neutrons and charged particles given by  
  
\begin{equation} 
t_{drag} \sim~4.5~\times 10^{-1}~ L_5^2~ T_8^2 ~ yr, 
\end{equation} 
where $L_5\equiv L/(10^5~cm)$ and $T_8\equiv T/(10^8 K).$ 
 
In the opposite regime, when the collisions between charged particles and neutrons are frequent, and the weak interaction rate is high, the relative diffusion of particles will be impeded by the collisional drag, but the system reaches the chemical quasi-equilibrium in a timescale controlled by weak interactions of magnitude (assuming modified Urca reactions) 
 
\begin{equation}  
t_{weak} \sim 4.3 ~ \times 10^5~ T_8^{-6} ~yr. 
\end{equation} 
 
\subsubsection{Timescale for magnetic field evolution}  
\label{conc13} 
The Ohmic dissipation promotes the decay of the magnetic field in a timescale 
 
\begin{equation} 
t_{ohmic}~\sim 1.4~\times~ 10^{11}~ L_5^{2}~ T_8^{-2}~yr. 
\end{equation} 

This extremely long time implies that Ohmic dissipation is not very effective in producing magnetic-field decay in neutron stars. A more rapid decay of the magnetic field can occur through ambipolar diffusion, i.e., a joint drift of the magnetic field and the charged particles relative to the neutrons, in which magnetic energy becomes dissipated by collisions.  This diffusion process induces local density perturbations that deviate from chemical quasi-equilibrium. Weak interactions tend to erase the chemical imbalance by converting particles of one species into the other. During these conversions, neutrinos, and antineutrinos remove part of the magnetic energy. In estimating the timescale of magnetic field decay induced by these processes, it is useful to distinguish between the two regimes we discussed above.  
 
If ambipolar diffusion occurs far more rapidly than the weak interaction process, i.e.,  $t_{drag}\ll t_{weak},$ diffusive quasi-equilibrium is quickly reached, but chemical quasi-equilibrium is not. Thus, the rate at which chemical quasi-equilibrium is restored by weak interactions determines the characteristic timescale over which the magnetic field evolves,  
 
\begin{equation} 
\label{tbb1} 
t_B~\sim~ 1.7\times~ 10^{9}~ B_{15}^{-2}~T_8^{-6}~yr. 
\end{equation} 
This timescale is the corresponding one to reach chemical quasi-equilibrium $t_{weak}$ but amplified by a factor of the order of the ratio of the charged fluid pressure to the magnetic pressure. This is expected since the Lorentz force drives the ambipolar diffusion that prevents weak interactions from restoring chemical quasi-equilibrium. 
 
In the opposite regime, collisions are frequent and weak interactions are fast, i.e. $t_{weak}\ll t_{drag},$ so the system reaches chemical quasi-equilibrium long before diffusive quasi-equilibrium. Therefore, the evolution of the magnetic field is limited by the collisions that control the ambipolar diffusion, on a timescale of   
 
\begin{equation} 
\label{tbb2} 
t_B~\sim~ 1.8\times~ 10^{3}~ B_{15}^{-2}~ L_{5}^2~ T_8^{2}~yr.  
\end{equation} 
Since the magnetic field sustains the diffusive imbalance and drives the diffusion, the timescale given in Eq.~(\ref{tbb2}) is similar to that taken by particles to achieve diffusive quasi-equilibrium but, as above, amplified by a factor that depends on the magnetic field strength.  
 
In the limit of non-interacting protons and neutrons studied by GR-92, Eqs.~(\ref{tbb1}) and (\ref{tbb2}) correspond to the estimate of the timescale for magnetic field evolution due to the ambipolar diffusion process of Eq.~(35) in GR-92. We note that GR-92 considered the neutrons to be a static species in diffusive quasi-equilibrium. From this correspondence, we confirm that the motion of the neutrons, which was included in our model, plays no important role in the one-dimensional evolution. 
 
\subsection{Normal-mode solutions}
\label{conc2} 
 
In the spirit of finding  an analytical solution to the set of non-linear partial differential equations that describes the evolution of the system, we completed a linear perturbation analysis of the equations in the limit of a nearly uniform system and tried a normal mode solution (Sect.~\ref{normmodes}). We found three exponentially-decaying modes that could be interpreted as the episodes of successive relaxation to the quasi-equilibrium states identified in the previous physical analysis (see  Sects.~\ref{carsca} and ~\ref{conc1}). We found analytical and numerical solutions for the decay times of these modes (see Appendix \ref{ApA}) that agree with the estimated formulae for the evolutionary timescales of the general non-linear system (see  Sects.~\ref{carsca}, \ref{conc1} and Figs.~\ref{t1},\ref{t2}, and \ref{t3}). 
 
\subsection{Numerical integration of the partial differential equations}
\label{conc3} 
 
We constructed a finite-difference numerical code to solve the full system of non-linear partial differential equations. This was tested against the normal-mode solutions (Fig.~ \ref{linen}) and by verifying the known conservation laws of magnetic flux and baryon number in highly non-linear situations (Fig.~ \ref{gauss}). Applying this numerical code, we propose to consider, in the future, conditions in which the background is non-homogeneous, as well as non-linear situations. 
 
\subsection{Conclusions}
\label{summa} 
 
We have established a general multifluid formalism in which it is possible to study the magnetic field decay processes in neutron stars identified by GR-92, which are likely to represent the basis of the magnetar phenomenon. As a first step, we have focused on a simplified geometry in which the magnetic field points in one Cartesian direction and varies in an orthogonal direction to it. We estimated the timescales of the relevant processes, and using numerical simulations we followed the temporal evolution of some simple magnetic-field configurations. The present work is far from exhausting the possibilities of this formalism, which should be applied to more realistic situations, including neutron stars with spherical symmetry, non-uniform background density and three-dimensional magnetic fields with components that depend on two or three spatial coordinates. Further steps might consider convective motions and additional species of particles as well as the effects of superfluidity and superconductivity. 
 
\section{Appendix} 
 
\subsection{Solution to the linear system}
\label{ApA} 
 
The set of Eqs.~(\ref{magbar}), (\ref{nbnorm}), and (\ref{ncnorm}) is, in general, non-linear with respect to the magnetic field variable $B_z$. To find a linear solution to this set, we linearized the equations respect to the magnetic field, as described in Sec.~\ref{normmodes}. Assuming that the properties of the background star are homogeneous, that is quantities with sub-index zero do not depend on position, we obtain the following linear set of equations: 
 
\begin{equation} 
\label{setlin} 
\frac{\partial \bf {\overline{Q}}}{\partial \overline{t}}=\overline{\bf{D}}_0\frac{\partial ^2\overline{\bf{Q}}}{\partial \overline{x}^2}+\overline{\bf{W}}_0~\overline{\bf{Q}},\end{equation} 
where we define the solution vector to be     
 
\begin{equation} 
\label{solvect} 
\bf\overline{Q}= \left(\begin{array}{c} 
\delta\overline{n}_B\\ 
\delta \overline{n}_c \\ 
\delta \overline{B}_z 
\end{array}\right), 
\end{equation} 
 
\noindent with the coupling matrices 
 
\begin{equation} 
{\overline{\bf{D}}_0}= 
\left(\begin{array}{ccc} 
d_{11}&d_{12}&d_{13}\\ 
d_{21}&d_{22}&d_{23}\\ 
d_{31}&d_{32}&d_{33}\\ 
\end{array}\right), 
\end{equation} 
where
\begin{equation}
d_{11}=(1+n)[1+\beta(1+\frac{\epsilon}{1+n})], 
\end{equation}

\begin{equation}
d_{12}=k(1+n)[1+\frac{\varrho}{k}(1+\frac{\epsilon}{1+n})],
\end{equation}

\begin{equation}
d_{13}=2(1+n)[1+\frac{\epsilon}{1+n}],
\end{equation}

\begin{equation}
d_{21}=n[1+\beta(1+\frac{\epsilon}{n})],
\end{equation}

\begin{equation}
d_{22}=kn[1+\frac{\varrho}{k}(1+\frac{\epsilon}{n})],
\end{equation}

\begin{equation}
d_{23}=2n[1+\frac{\epsilon}{n}],
\end{equation}

\begin{equation}
d_{31}=\Upsilon d_{21}, \quad d_{32}=\Upsilon d_{22}, \quad d_{33}=\omega+\Upsilon d_{23}
\end{equation}

and 

\begin{equation} 
\overline{\bf{W}}_0= 
\left(\begin{array}{ccc} 
w_{11}&w_{12}&w_{13}\\ 
w_{21}&w_{22}&w_{23}\\ 
w_{31}&w_{32}&w_{33}\\ 
\end{array}\right), 
\end{equation} 
where
\begin{equation}
w_{11}=w_{12}=w_{13}=w_{23}=w_{31}=w_{32}=w_{33}=0,
\end{equation}

\begin{equation}
w_{21}=-n\theta(\frac{\beta}{n}-1),
\end{equation}

\begin{equation}
w_{22}=-n\theta(\frac{\varrho}{n}-k).
\end{equation}
To simplify the notation, we define $\overline \omega \equiv \omega$, $\overline \epsilon \equiv \epsilon$, $\overline \Upsilon \equiv \Upsilon$, $\overline \theta \equiv \theta$, $\overline{r}_0 \equiv r$, $\overline{n}_{0c}\equiv n$, $\overline{k}_{nc}\equiv k$, $\varrho\equiv n\overline{k}_{cc}$, $\beta\equiv n\overline{k}_{pB}$. We find normal-mode solutions to this linear system of the form 
 
\begin{equation} 
\label{norm1} 
\overline{\bf {Q}}_m=\exp(-\overline{\eta}_m\overline t)\cos(l\pi\overline{x})\overline{\bf{A}}_m, 
\end{equation} 
where $m=1,2,3$ labels each mode, and we order the decay times in increasing order. The decay time for each mode is  
 
\begin{equation} 
\label{eigentimes} 
\overline \tau_m=1/ \overline \eta_m,  
\end{equation} 
and the amplitudes are 
 
\begin{equation} 
\label{eigenvec} 
\overline{{\bf A}}_m= \left(\begin{array}{c} 
\overline{a}_{\delta \overline{n}_B}(\overline{\eta}_m)\\ 
\overline{a}_{\delta \overline{n}_c}(\overline{\eta}_m)\\ 
\overline{a}_{\delta \overline{B}_z}(\overline{\eta}_m)\\ 
\end{array}\right). 
\end{equation} 
where $\overline{a}_{\delta \overline{B}_z}$ is normalized with respect to $B_c$, while $\overline{a}_{\delta \overline{n}_{B,c}}$ are both normalized with respect to $\delta n_s=B_c^2/(8\pi n_{0n} k_{nB})$. If we substitute Eq.~(\ref{norm1}) into the linear matrix system Eq.~(\ref{setlin}), we obtain the eigenvalue problem 
 
\begin{equation} 
\label{eigen} 
\overline{{\bf C}}_0 \overline{{\bf A}}_m = \overline{\eta}_m \overline{{\bf A}}_m, 
\end{equation} 
that needs to be solved, where 
\begin{equation} 
\label{matrix} 
\overline{{\bf C}}_0=(l\pi)^2\overline{{\bf D}}_0-\overline{{\bf W}}_0. 
\end{equation}

\subsection{Analytical solution for the decay times} 
\label{ApB} 
 
To obtain analytical expressions for the decay times of the normal modes [see Eq.~(\ref{eigentimes})], we write the matrix of Eq.~(\ref{matrix}) as

\begin{equation} 
\label{matrsm} 
{\bf\overline{C_0}}= 
\left(\begin{array}{ccc} 
c_{11}&c_{12}&c_{13}\\ 
uc_{11}+\chi_{21}&uc_{12}+\chi_{22}&uc_{13}+\chi_{23}\\ 
fc_{11}+\chi_{31}&fc_{12}+\chi_{32}&fc_{13}+\chi_{33}\\ 
\end{array}\right), 
\end{equation} 
where we define $n^*=1+n$, $u=n/n^*$, $\epsilon^*=\epsilon/n^*$, $f=u\Upsilon$, and 
 
\begin{equation} 
c_{11}=(l\pi)^2 n^*\left(1+\beta\left(1+\epsilon^*\right)\right), 
\end{equation} 
 
\begin{equation} 
c_{12}=(l\pi)^2 kn^*\left(1+\frac{\rho}{k}\left(1+\epsilon^*\right)\right), 
\end{equation} 
 
\begin{equation} 
c_{13}=2n^*(l\pi)^2\left(1+\epsilon^*\right),  
\end{equation} 
 
\begin{equation} 
\chi_{21}=(l\pi)^2\beta \epsilon^*+n \theta\left(\frac{\beta}{n}-1\right),  
\end{equation} 

\begin{equation} 
\chi_{22}=(l\pi)^2\rho \epsilon^*+n \theta\left(\frac{\rho}{n}-k\right), 
\end{equation} 
 
\begin{equation} 
\chi_{23}=2 (l\pi)^2\epsilon^*, \quad \chi_{31}=\frac{\Upsilon \beta \chi_{23}}{2},
\end{equation}

\begin{equation} 
\chi_{32}=\frac{\Upsilon \rho \chi_{23}}{2}, \quad \chi_{33}=\Upsilon \chi_{23}+(l\pi)^2\omega.
\end{equation}
We also note that all $\chi_{ij}\ll 1$ whereas $c_{ij}\sim 1$. In the limit $\chi_{ij}=0$, the eigenvalues are $\eta_1=c_{11}+uc_{12}+fc_{13}$, $\eta_2=\eta_3=0$. In the general case, we can write 
\begin{eqnarray} 
\eta_1&=&c_{11}+uc_{12}+fc_{13}+\varphi\nonumber\\
&&=(l\pi)^2n^*[1+\beta+n(2\Upsilon +k)\nonumber\\
&&+\epsilon^*(\beta+n(2\Upsilon+\rho))]+\varphi, 
\end{eqnarray} 
and treat $\eta_2,\eta_3,\varphi \sim O(\chi_{ij})$ as small quantities, which we derive below. We calculate the characteristic polynomial of the matrix Eq.~(\ref{matrsm}) and compare terms of the same order in this polynomial with those in the equation 
 
\begin{equation} 
(\eta_1-\eta)(\eta_2-\eta)(\eta_3-\eta)=0. 
\end{equation} 
After neglecting quadratic and cubic terms in the small quantities $\eta_2,\eta_3, and\varphi$, we find 
 
\begin{equation} 
\eta_2+\eta_3=F, 
\end{equation}  
 
\begin{equation} 
\eta_2\eta_3=G, 
\end{equation}  
 
\begin{equation} 
\varphi=(\chi_{22}+\chi_{33})-F, 
\end{equation}  
where 

\begin{equation} 
F=\frac{c_{11}F_1-c_{12}F_2-c_{13}F_3}{(l\pi)^2n^*[1+\beta+n(2\Upsilon +k)+\epsilon^*(\beta+n(2\Upsilon+\rho))]}, 
\end{equation} 
with the definitions,

\begin{equation}
F_{1}=\chi_{22}+\chi_{23},
\end{equation}

\begin{equation}
F_{2}=\chi_{21}+f\chi_{23}-u\chi_{33},
\end{equation}

\begin{equation}
F_{3}=u\chi_{32}+\chi_{31}-f\chi_{22},
\end{equation}

and

\begin{equation} 
G=\frac{c_{11}G_1+c_{12}G_2+c_{13}G_3}{(l\pi)^2n^*[1+\beta-n(2\Upsilon +k)+\epsilon^*(\beta-n(2\Upsilon+\rho))]}, 
\end{equation}

with,

\begin{equation}
G_{1}=\chi_{22}\chi_{33}-\chi_{23}\chi_{32},
\end{equation}

\begin{equation}
G_{2}=\chi_{23}\chi_{31}-\chi_{21}\chi_{33},
\end{equation}

\begin{equation}
G_{3}=\chi_{21}\chi_{32}-\chi_{22}\chi_{31}.
\end{equation}

Thus, the eigenvalues are given by 
 
\begin{equation} 
\label{val1} 
\eta_1=(l\pi)^2n^*[1+\beta+n(2\Upsilon +k)+\epsilon^*(\beta+n(2\Upsilon+\rho))]+\varphi, 
\end{equation} 
 
\begin{equation} 
\label{val2val3} 
\eta_2,\eta_3=\frac{F\pm \sqrt{F^2-4G}}{2}. 
\end{equation} 
Equations. ~(\ref{val1}) and (\ref{val2val3}) give approximate analytical expressions for the decay times of the normal modes [see Eq.~(\ref{eigentimes})] as functions of the dimensionless parameters, $\overline \tau_m(\epsilon,\Upsilon, \theta, \omega)$. Equations.~(\ref{val1}) and (\ref{val2val3}) are useful because from them we can understand in an analytical way the relative importance of the different physical processes for the behavior of the decay times. The linear decay times $\overline{\tau}_m$ correspond to the estimated characteristic timescales $\overline{t}_{m}$ in the general non-linear regime (see Eqs. ~(\ref{torder1}), (\ref{torder2}), and (\ref{torder3})). 
 
\subsection{Numerical Method} 
 
To solve the dimensionless system of Eqs.~(\ref{magbar}), (\ref{nbnorm}), and (\ref{ncnorm}) numerically we discretize them using an FTCS (Forward Time Centered Space) first order scheme. We define the spatial and temporal grid by $\overline{x}_j=j\Delta \overline{x}$, $\overline{t}_n=n\Delta \overline{t}$ where $j=0,1,2,...,j_{max}$ and $n=0,1,2,...,n_{max}$. In this method, we evaluate the spatial derivatives of a quantity $A(\overline{x},\overline{t})$ at a half point between consecutive grid points, namely,  
 
\begin{equation} 
\left(\frac {\partial A}{\partial \overline{x}}\right)_{j}=\frac{A_{j+1/2}^n-A_{j-1/2}^n}{\Delta \overline{x}}, 
\end{equation} 
 
\begin{equation} 
\left(\frac {\partial A}{\partial \overline{x}}\right)_{j+1/2}=\frac{A_{j+1}^n-A_j^n}{\Delta \overline{x}}, 
\end{equation} 
 
\begin{equation} 
\left(\frac {\partial A}{\partial \overline{x}}\right)_{j-1/2}=\frac{A_{j}^n-A_{j-1}^n}{\Delta \overline{x}}. 
\end{equation} 
Then we interpolate it to the grid using the known grid quantities, 
 
\begin{equation} 
A_{j+1/2}^n=\frac{A_{j+1}^n+A_j^n}{2},
\end{equation} 
 
\begin{equation} 
A_{j-1/2}^n=\frac{A_{j-1}^n+A_j^n}{2}. 
\end{equation} 
 
\begin{center}
{\bf Acknowledgements}
\end{center}

We thank FONDECYT for financial support through the postdoctoral project 3060103 and regular projects 1060644 and 1070854. We also thank the Gemini Project 32070014, the ESO-Chile Mixed Committee, and the FONDAP Center for Astrophysics (15010003). J.H. and A.R. are grateful for the hospitality of the Max Planck Institute for Astrophysics and the Max Planck Institute for Extraterrestrial Physics, in Garching, Germany, where part of this research was developed. We thank the Editor of {\it Astronomy}\&{\it Astrophysics} Steven Shore and an anonymous referee, whose suggestions and comments were useful to improve the manuscript.


\begin{thebibliography}{32}
\bibitem{B-95}  
  Bhattacharya D., 1995, in X-Ray Binaries, ed. W. H. G. Lewin, J. van Paradijs, \& E. P. J. van den Heuvel (Cambridge: Cambridge University Press), 233 
\bibitem{CZB-01}  
  Cumming, A., Zweibel, E., Bildsten, L. 2001, ApJ, 557, 958 
\bibitem{C-02} 
  Cumming, A. 2002, MNRAS, 333, 589 
\bibitem{PM-04}  
  Payne, D. J. B. \& Melatos, A. 2004, MNRAS, 351, 569 
\bibitem{DT-92}  
  Duncan, R.C., \& Thompson, C. 1992, ApJ, 392, L9 
\bibitem{TD-96}  
  Thompson, C. \& Duncan, R.C., 1996 ApJ, 473, 322 
\bibitem{ACT-04}  
  Arras, P., Cumming, A., \& Thompson, C. 2004, ApJ, 608, L49 
\bibitem{GR-92}  
  Goldreich, P., \& Reisenegger, A. 1992, ApJ, 395, 250 
\bibitem{J-88}  
  Jones, P. B. 1988, MNRAS, 233, 875 
\bibitem{RG-02}  
  Rheinhardt, M.,\& Geppert, U. 2002, Phys. Rev. Lett., 88, 101103 
\bibitem{GR-02}  
  Geppert, U., \& Rheinhardt, M. 2002, A\&A, 392, 1015 
\bibitem{HR-02}  
  Hollerbach, R., \& R\"{u}diger, G. 2002, MNRAS, 337, 216 
\bibitem{HR-04}  
  Hollerbach, R., \& R\"{u}diger, G. 2004, MNRAS, 347, 1273 
\bibitem{RKG-04} 
  Rheinhardt, M., Konenkov, D., \& Geppert, U. 2004, A\&A, 420, L33 
\bibitem{CAZ-04}  
  Cumming, A., Arras, P., \& Zweibel, E.G. 2004, ApJ, 609, 999 
\bibitem{R-05}  
  Reisenegger, A., Prieto, J.P., Benguria, R., Lai, D., \& Araya, P.A: 2005, in Magnetic Fields in the Universe: From Laboratory and Stars to Primordial Structures, AIP Conference Proceedings, vol. 784, eds. E.M. de Gouveia dal Pino, G. Lugones, \& A. Lazarian, p. 263 
\bibitem{R-07}  
  Reisenegger, A., Benguria, R., Prieto, J.P., Araya, P.A., \&  Lai, D.  2007, A\&A, 472, 233 
\bibitem{Rei-07}  
  Reisenegger, A., 2007, AN, 328, 1173 
\bibitem{Bay-69}  
  Baym, G., Pethick, C., \& Pines, D. 1969, Nature, 224, 872
\bibitem{BS-04}  
  Braithwaite, J., \& Spruit, H.C. 2004 , Nature, 431, 819 
\bibitem{BS-06}  
  Braithwaite, J., \& Spruit, H.C. 2006 , A\&A, 450, 1097 
\bibitem{BN-06}  
  Braithwaite, J., \& Nordlund, A. 2006, A\&A, 450, 1077 
\bibitem{B-08}  
  Braithwaite, J. 2008, MNRAS, 386, 1947  
\bibitem{A-98}  
  Akmal, A., Pandharipande, V.R. \&  Ravenhall, D.G. 1998, Phys. Rev. C, 58, 1804 
\bibitem{Rei-95}  
  Reisenegger, A., 1995, ApJ, 442, 749 
\bibitem{F-05}  
  Fern\'andez, R., \& Reisenegger, A. 2005, ApJ, 625, 291 
\bibitem{YS-90}  
  Yakovlev, D.G., \& Shalybkov, D.A., 1990, Soviet Astron.Lett., 16, 86 
\bibitem{S-89}  
  Sawyer, R.F. 1989, Phys. Rev. D., 39, 3804 
\bibitem{S-83} 
  Shapiro, S.L., \& Teukolsky, S.A. 1983, Black Holes, White Dwarfs, and Neutron Stars (New York:Wiley) 
\bibitem{HUY-90} 
  Haensel, P., Urpin, V.A., \& Yakovlev, D.G. 1990, A\&A, 229, 133 

\end{thebibliography}
\end{document}